\documentclass{iopart}
\usepackage{graphicx}
\graphicspath{{Images/}}
\usepackage{subfigure}
\usepackage{cite}
\usepackage{url}

\expandafter\let\csname equation*\endcsname\relax

\expandafter\let\csname endequation*\endcsname\relax
\usepackage{physics}
\usepackage{mathtools}
\usepackage{amsfonts}

\begin{document}
\title[LMT interferometer to measure the effect of gravity on positronium]{A large-momentum-transfer matter-wave interferometer to measure the effect of gravity on positronium}
\author{G Vinelli$^{1,2,3}$, F Castelli$^{4,5}$, R Ferragut$^{5,6}$, M Romé$^{4,5}$, M Sacerdoti$^{6}$, L Salvi$^{1,2,3}$, V Toso$^{4,5}$, M Giammarchi$^{5}$, G Rosi$^{1,2}$ and G M Tino$^{1,2,3}$}

\address{$^1$ Department of Physics and Astronomy, Università di Firenze, Via Sansone 1, 50019 Sesto Fiorentino, Italy}
\address{$^2$ INFN, Sezione di Firenze, Via Sansone 1, 50019 Sesto Fiorentino, Italy}
\address{$^3$ LENS, Via Nello Carrara 1, 50019 Sesto Fiorentino, Italy}
\address{$^4$ Department of Physics, Università degli Studi di Milano, Via Celoria 16, 20133 Milano, Italy}
\address{$^5$ INFN, Sezione di Milano, Via Celoria 16, 20133 Milano, Italy}
\address{$^6$ L-NESS and Department of Physiscs, Politecnico di Milano, Via Anzani 42, 22100 Como, Italy}

\ead{giuseppe.vinelli@unifi.it}

\begin{abstract}
This paper reports the study of a new interferometric configuration to measure the effect of gravity on positronium. A Mach-Zehnder matter-wave interferometer has been designed to operate with single-photon transitions and to transfer high momentum to a 200 eV positronium beam. The work shows the results and methods used to simulate the interferometer and estimate the operating parameters and the time needed to perform the experiment. It has been estimated that within less than one year, the acquisition time is sufficient to achieve a 10\% accuracy level in measuring positronium gravitational acceleration, even with a poorly collimated beam, which is significant for theoretical models describing matter-antimatter symmetry. These results pave the way for single photon transition large momentum transfer interferometry with fast atomic beams, which is particularly useful for studies with antimatter and unstable atoms.
\end{abstract}

\submitto{\CQG}
\maketitle

\section{Introduction}\label{Sec:Intro}
Positronium (Ps) is the hydrogen-like quasi-stable bound system of an electron and its antiparticle, the positron. Ps is the lightest element (about 10$^3$ times lighter than hydrogen) and exists in the ground state in two sublevels depending on the spins of the electron and positron: singlet (para-Ps, spin 0) and triplet (ortho-Ps, spin 1). Given the multiplicity of these states, positronium in the ground state forms as 75\% ortho-Ps and 25\% para-Ps. Their lifetimes in vacuum are very different, 0.125 and 142 ns for para-Ps and ortho-Ps, respectively. Also, the annihilation characteristics in vacuum are different: para-Ps annihilates with emission of two gamma rays of 511 keV each (equal to $m_e$c$^2$, with $m_e$ the electron rest mass), while ortho-Ps annihilates by emitting at least three gamma rays with a maximum energy of 511 keV each and with a total energy of $2m_e$c$^2$. Given the short lifetime of para-Ps, only the triplet state (ortho-Ps) is of interest for this work and in the context of this paper Ps will always mean ortho-Ps.

Since its discovery \cite{PhysRev.82.455}, Ps has been the subject of many experimental and theoretical investigations (see e.g. \cite{RevModPhys.53.127,LIANG1988419}).
Indeed, as the simplest purely leptonic bound state, Ps, together with muonium, gives a privileged opportunity for high-precision studies of quantum electrodynamics (QED) \cite{KARSHENBOIM2004,KARSHENBOIM20051}: the low electron and positron mass implies that weak force contributions to Ps energy levels are negligible \cite{GOVAERTS1996451}, and that Ps is well described as a bound-state QED system. Measured deviations from theoretical predictions can then be interpreted as possible indications of new physics \cite{RevModPhys.53.127,Castelli2012,PhysRevA.40.5526,PhysRevA.7.447,PhysRevA.8.625}.
Being neutrally charged, Ps has also the advantage of being insensitive to external electric fields that probably represent the main difficulty in measuring the gravitational fall of free electrons \cite{e-Fall}.
In recent decades, interest in this atom has grown considerably also thanks to the development of positronium beam technology \cite{Cassidy2018}.
Two of the main limitations in dealing with the Ps atom are its reduced lifetime and the possibility of annihilation with electrons in the surrounding environment; this implies that to obtain intense Ps beams it is necessary to work with high Ps velocities and efficiently guide the atoms in an ultra-high vacuum system.

The effect of the Earth's gravitational field on antimatter is the subject of ongoing research and discussions at present and during the last decades \cite{RevModPhys.29.423,doi:10.1119/1.1996159,SCHERK1979265,NIETO1991221,PhysRevLett.66.850,HUBER20001245,Chardin2018,Rousselle2022,OBERTHALER2002129,Mariazzi2020,Zimmer21,Cassidy2014}: any differences in the gravitational behavior between matter and antimatter would in fact represent a violation of Einstein Equivalence Principle (EEP). Antimatter gravitation has never been measured, except for a result obtained by the ALPHA collaboration at CERN, leading to the exclusion of a ratio between gravitational and inertial mass of anti-hydrogen greater than about 75 \cite{Amole2013}, not enough to address theoretical issues about fundamental laws.

This kind of experiments can often take advantage of the accuracy of interferometry to measure gravitational effects. Solutions based on different types of interferometry have been proposed, also for antimatter \cite{PhysRevLett.129.173204}. Some of these rely on Ps excitation to high-order Rydberg states to increase its lifetime \cite{MILLS2002102,PhysRevA.94.012507}.

In this paper a new way of measuring the gravitational effect on the Ps atom using a Large Momentum Transfer (LMT) Mach-Zehnder interferometer is proposed.  We discuss the geometry, show numerical simulation results, and estimate the data taking time required to achieve a given accuracy. The article is organized as follows. After a presentation of the theoretical interest in Ps gravitation in section \ref{Sec:PosGrav}, we describe in section \ref{Sec:LMTI} the LMT interferometer, its main operating parameters, the experimental setup and the proposed noise rejection method. In section \ref{Sec:Sim} we describe the numerical simulations and compare a full quantum mechanical approach with a semi-classical Monte Carlo computation. Moreover, in section \ref{Sec:Res} we discuss the simulation results that allowed us to define the design parameters of the interferometer, the detection strategy and estimate the acquisition time of the experiment. The results are compared with another interferometric light-pulses approach proposed in the literature \cite{OBERTHALER2002129} in section \ref{Sec:Com}.

\section{Positronium gravitation}\label{Sec:PosGrav}
The study of gravitation with systems containing antimatter addresses some of the main unresolved issues in Modern Physics. 

First of all, our Universe features a striking particle-antiparticle asymmetry; according to the widely accepted Big Bang scenario \cite{Mukhanov}, the unbalance was generated close to the Grand Unification era, around 10$^{-35}$ s of cosmic evolution time. Therefore,  at the time of baryon formation (temperature of 10$^{13}$ K, or 10$^{-6}$ s) only protons and neutrons could be formed, and not their antiparticles. The likely explanation of this effect is some level of violation of fundamental laws (like CP invariance, the combination of Charge and Parity symmetries) as well as the occurrence of the Sakharov conditions \cite{Sakharov}, all of this implying that only 
baryons (and not anti-baryons) fueled the first stages of cosmic nucleosynthesis.

While the Standard Model of particle physics has the possibility of naturally generating CP violation for instance through the phase of the Cabibbo-Kobayashi-Maskawa mixing matrix in the hadronic sector, the amount of violation seems largely insufficient to explain the primordial asymmetry. 

The Standard Model itself, despite of its remarkable successes, appears to be an incomplete theory for a variety of reasons: it does not include gravitational interaction, and does not explain the oscillation of neutrinos or the existence of dark matter and dark energy, constituting the vast majority of the energy budget of the Universe. Finally, it does not satisfactorily explain the predominance of matter over antimatter in the Universe. Moreover, it seems natural that new Physics or some violation of fundamental laws would appear at energies close to the Planck-scale, because of the unavoidable interplay between Quantum Physics and Gravitation, which in turn could lead to residual effects at the present energy level.

For these reasons, the study of antimatter holds the possibility of both improving the Standard Model and to shed light on the composition of our Universe. Matter and antimatter are related by two key symmetries of Modern Physics: Charge-Parity-Time reversal (CPT) and the Einstein Equivalence Principle (EEP). CPT invariance is the main theoretical tool linking particles and antiparticles at the quantum level and in a flat spacetime. The EEP, on the other hand, relates any form of matter and energy, for instance, through the feature of Universal Free Fall, but only at the macroscopic level in any curved spacetime. For these reasons matter and antimatter asymmetry can effectively be addressed by studying the behaviour of subatomic particles in a gravitational field.  

In spite of its short lifetime, Ps offers two important features which make it a very interesting system from this point of view. First of all, it is the only bound system made of elementary constituents of the Standard Model. Its mass is therefore directly composed by key parameters of the model, while (for instance) the mass of an antiproton is mostly made of the energy of the color field, and not of the mass of the quarks. 
From the theoretical viewpoint, possible violations of fundamental symmetries (Lorentz Invariance, CPT and the EEP) are related by the Greenberg's theorem \cite{PhysRevLett.89.231602} and – more in general – by the framework of the Standard Model Extension (SME) \cite{PhysRevD.55.6760,PhysRevD.58.116002}. 
The SME \cite{PhysRevD.55.6760,PhysRevD.58.116002} is an effective field theory combining General Relativity and the Standard Model, as well as including in a gauge-invariant way all the possible spacetime operators violating CPT and Lorentz symmetry. It is then possible to parametrize the presence of these Lorentz and CPT violating operators, which constitute parameters of the model to be measured experimentally. For instance, in the frame of the SME, it has been shown that gravitation with Ps can address relevant physical parameters already at the 10\% accuracy level \cite{PhysRevD.92.056002}.

\section{LMT interferometer for inertial sensing with antimatter}\label{Sec:LMTI}
Atom interferometry is a versatile and powerful tool for precision measurements, in particular in gravitational physics \cite{RevModPhys.81.1051,APeters_2001,PhysRevA.65.033608,PhysRevLett.97.240801,doi:10.1126/science.aap7706,Rosi2014,Tino2021}. The light-pulse interferometer uses sequences of optical pulses to split, redirect and recombine matter waves by transferring photon momenta \cite{PhysRevLett.67.181,PhysRevLett.75.2638} and emulating optical elements (e.g., mirrors and beam splitters) for coherent manipulation of atomic wave packets \cite{BORDE198910}. This type of interferometer has been proposed to improve sensitivity to inertial forces with large momentum transfer techniques using several light pulses to increase the space-time area of the interferometer \cite{PhysRevLett.85.4498}. 

The type of LMT interferometer that we propose can represent an improvement with respect to other types of interferometers based on high-order Bragg processes \cite{PhysRevLett.100.180405}, Raman-Nath standing-wave interactions \cite{PhysRevLett.79.784} or adiabatic transfer \cite{PhysRevLett.73.2563,Wicht_2002} that impose too restrictive constraints on the angular spread of the atomic beam or are unable to populate higher-lying momentum states. Conventional light-pulse atom interferometry uses two-photon interactions. However, some challenging applications, such as ultralight dark matter searches \cite{PhysRevD.97.075020,El-Neaj2020} and gravitational wave detection \cite{PhysRevD.78.122002,Yu2011,PhysRevD.93.021101,Loriani_2019,doi:10.1142/S0218271819400054,Schubert2019}, prefer the use of single photon transitions. Multiple pulses LMT-enhanced interferometry is an active research field and new configurations based on single photon transitions have been recently proposed for atomic clocks \cite{PhysRevLett.124.083604} and as a method to reach the required sensitivity while keeping low levels of laser noise \cite{Graham2017}. Indeed, this kind of interferometer can be particularly convenient because it allows the use of phase noise suppression techniques such as the one proposed in this paper \cite{PhysRevLett.110.171102}.

Generically, the working principle of an interferometer for inertial sensing is based on the fact that an atom is split into a quantum superposition of states whose wave functions accumulate a phase shift proportional to the gravity acceleration. The relationship between the phase shift $\Delta\Phi$ and a generic gravity acceleration $a$ is
\begin{eqnarray}
    \Delta\Phi=k_{\rm eff}aT^2
    \label{eqn:Phase}
\end{eqnarray}
with $T$ the interrogation time (half of the propagation time through the whole interferometer) and $k_{\rm eff}$ the effective transferred momentum. The relevant quantity to be measured is the number of atoms leaving one of the two arms of the interferometer, which is proportional to the phase shift \cite{OBERTHALER2002129,BATELAAN199785}. \Eref{eqn:Phase} shows that the quantities to be maximized to obtain a large signal are $k_{\rm eff}$ and $T$.
Since Ps is unstable, designing an interferometer with too high T would lead to a severe reduction of the beam population. For this reason, we decided to maximize $k_{\rm eff}$ by designing a LMT interferometer with $k_{\rm eff} = 10k=10\frac{2\pi}{\lambda_{\rm L}}$, where $\lambda_{\rm L}$ is the wavelength of the laser.
Moreover, the need to avoid a rapid annihilation of the Ps leads us to consider one of the excited $n=2$ Ps states, precisely the 2$^3$S state, which has an annihilation lifetime of 1.13 $\mu$s and a radiative lifetime of 0.2 s. This state therefore guarantees a much higher survival probability of the Ps atoms with respect to the ground (0.142 $\mu$s lifetime) ortho-positronium state. 

In this work we present the design of a LMT Mach-Zehnder interferometer which makes use of single-photon transitions between the above state and the dipole-allowed $n=3$ Ps states.
For the proposed configuration, we have fixed the interferometer length to the experimentally feasible value of 5 m, and we have chosen the atomic beam speed equal to $6\times10^6$ m/s as a reasonable compromise between annihilation and interrogation time which was set at about 410 ns.

The laser light has 1312 nm wavelength and is circularly polarized to exploit the 2-3 transition by using the states 2$^3$S$_1$ with $m_{\rm J}$=1 and 3$^3$P$_2$ with $m_{\rm J}$=2. We have chosen these states and light polarization to ensure that the transition can be well described by a two-level system formalism. The Ps must therefore enter the interferometer in the state 2$^3$S$_1$ with $m_{\rm J}$=1. 

One way to obtain a Ps collimated beam is to guide Ps$^-$ ions leaving a positron-Ps$^-$ converter \cite{Nagashima_2006,NAGASHIMA201495,Michishio}. The advantage of producing Ps through the negative Ps$^-$ ion, compared to Ps formed in mesoporous materials \cite{Consolati2013,PhysRevA.98.013402,PhysRevA.81.052703}, is the possibility to obtain a tunable and monoenergetic beam with low divergence and high coherence well suited for interferometric experiments \cite{Nagashima_2021}. On the other hand, the disadvantage is the short Ps$^-$ lifetime (479 ps), so a very compact system is needed to produce, accelerate and focus the Ps$^-$ beam via electrodes. This topic is discussed in Ref. \cite{Sacerdoti}, where a proposal to produce a Ps beam is presented: the idea is based on production, acceleration and focusing of Ps$^-$ ions, followed by photodetachment of one electron via high-power cavity pumped by an Erbium fiber laser \cite{Igarashi_2000}, leaving Ps essentially in the ground triplet state.

In Ref. \cite{Sacerdoti} the authors propose to use a LINAC as primary electron beam to produce Ps. In Ref. \cite{Brixino,Drebot:2022nho,Bacci}, a superconductive LINAC operates at 92 MHz with an average electron current of 2.5 mA and an acceleration energy of 10 MeV, producing at least 10$^{16}$ fast electrons per second with a beam spot lower than 1 mm and a divergence of 50 µrad. Taking into account the e$^+$/e$^-$ conversion efficiency \cite{CHARLTON2021164657}, about 10$^{10}$ e$^+$/s are expected. Considering the Ps$^-$/e$^+$ \cite{NAGASHIMA201495} and the simulated Ps/Ps$^-$ conversion efficiencies, about 0.5x10$^8$ Ps/s are expected to enter the interferometer. In Ref. \cite{Sacerdoti} a method to reduce the atomic beam spatial dimension based on its remoderation that could increase the interferometric efficiency is also cited.

Photodetachment process has been simulated based on the literature \cite{Cooper1968,Michishio2019} and the results show that it has a destructive effect on the shape of the atomic beam, which reaches a maximum angular divergence of about 10 mrad and $\pm$10 eV as energy spread for a 1560 nm laser.
To reduce the effect of photodetachment, one can consider using a laser at a wavelength closer to the process threshold, such as 3600 nm. To compensate for the lowering of the cross section, however, it would be necessary to increase the power by about an order of magnitude, which would probably require a pulsed system.
Under these conditions the divergence of the atomic beam is little altered by the process and depends mainly on the generation and focusing system upstream of the photodetachment  \cite{Sacerdoti}. However, since we also propose to use it for the ionization stage and to simplify the design of the experiment, we will refer to the 1560 nm laser system, which is the most pessimistic case.

After the Ps beam production stage, the atoms will be prepared according to the following additional steps:
\begin{itemize}
\renewcommand\labelitemi{}
  \item i.   243 nm laser excitation to 2$^3$P$_0$;
  \item ii. 18 GHz microwave excitation to 2$^3$S$_1$ ($m_{\rm J}$=1).
\end{itemize}

\begin{figure}[ht]
    \centering
    \includegraphics[width=1\textwidth]{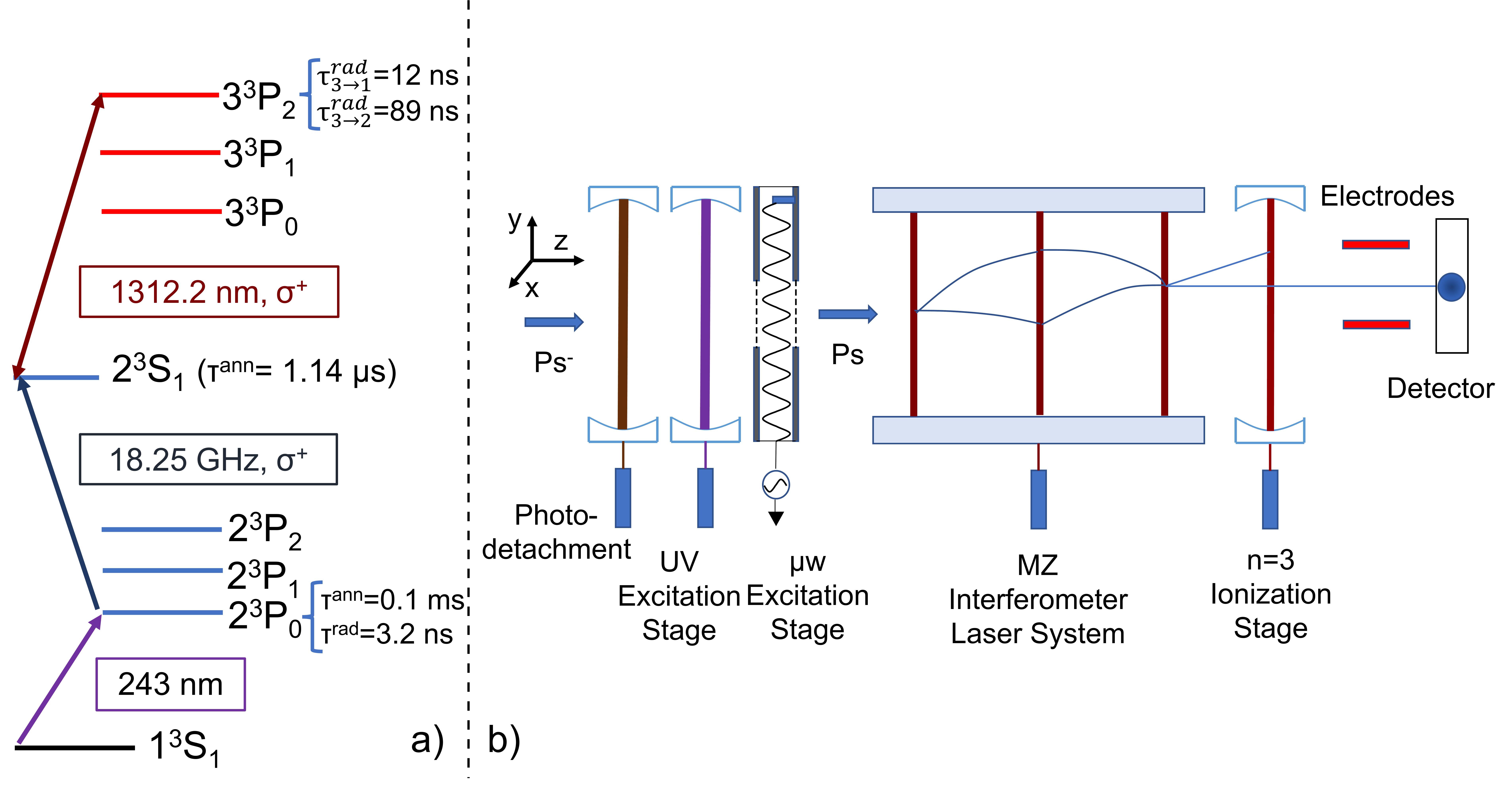}
    \caption{a) Energy level scheme of Ps, transitions and lifetimes \cite{PhysRevA.100.063414, Cassidy2018}. b) Representation of the Ps preparation, interferometric and detection stages. The Ps$^-$ beam is photodetached so that the resulting Ps beam is excited to the 2$^3$S$_1$ state and propagates through the Mach-Zehnder interferometer. At the end of the apparatus, the $n=3$ state is ionized, and the particles that have remained charged are swept away by a moderate electric field, leaving the Ps ($n=2$) state to be detected. The length of the interferometer is 5 m. In the scheme, gravity acts along the negative y direction.}
    \label{fig:setup}
\end{figure}

For the UV transition, a magnetic field is needed to define a quantization axis. To avoid Doppler effects in excitation the laser beam is directed perpendicularly to the Ps collimated beam. Excitation of one or all the ground triplet states, corresponding to transitions $\Delta m_{\rm J}$=0, $\pm1$, is obtained with suitable linearly polarized light.
The second excitation uses circular polarization to selectively couple the 2$^3$P$_0$ and 2$^3$S$_1$ in a $\sigma +$ transition. A similar UV-microwave excitation scheme can be found in Ref. \cite{PhysRevLett.125.073002}. Starting from Ps$^-$, the Ps preparation, the interferometer and the detection stage are shown in \fref{fig:setup}.
 Note that the state chosen for the interferferometer (2$^3$S$_1$ with $m_{\rm J}$=1) is optimized to be as insensitive as possible to external electric and magnetic fields: $m_{\rm s}$=$\pm$1 states with zero azimuthal quantum number experience a negligible effect of magnetic fields \cite{PhysRevA.41.3478}, and electric fields have only second-order effects for $m_{\rm J}$=1 \cite{Cassidy2018}. This makes the system particularly robust with respect to external perturbations.
The Ps preparation stage is not the focus of this work so it will not be covered further.

\begin{figure}[ht]
    \centering
    \includegraphics[width=1\textwidth]{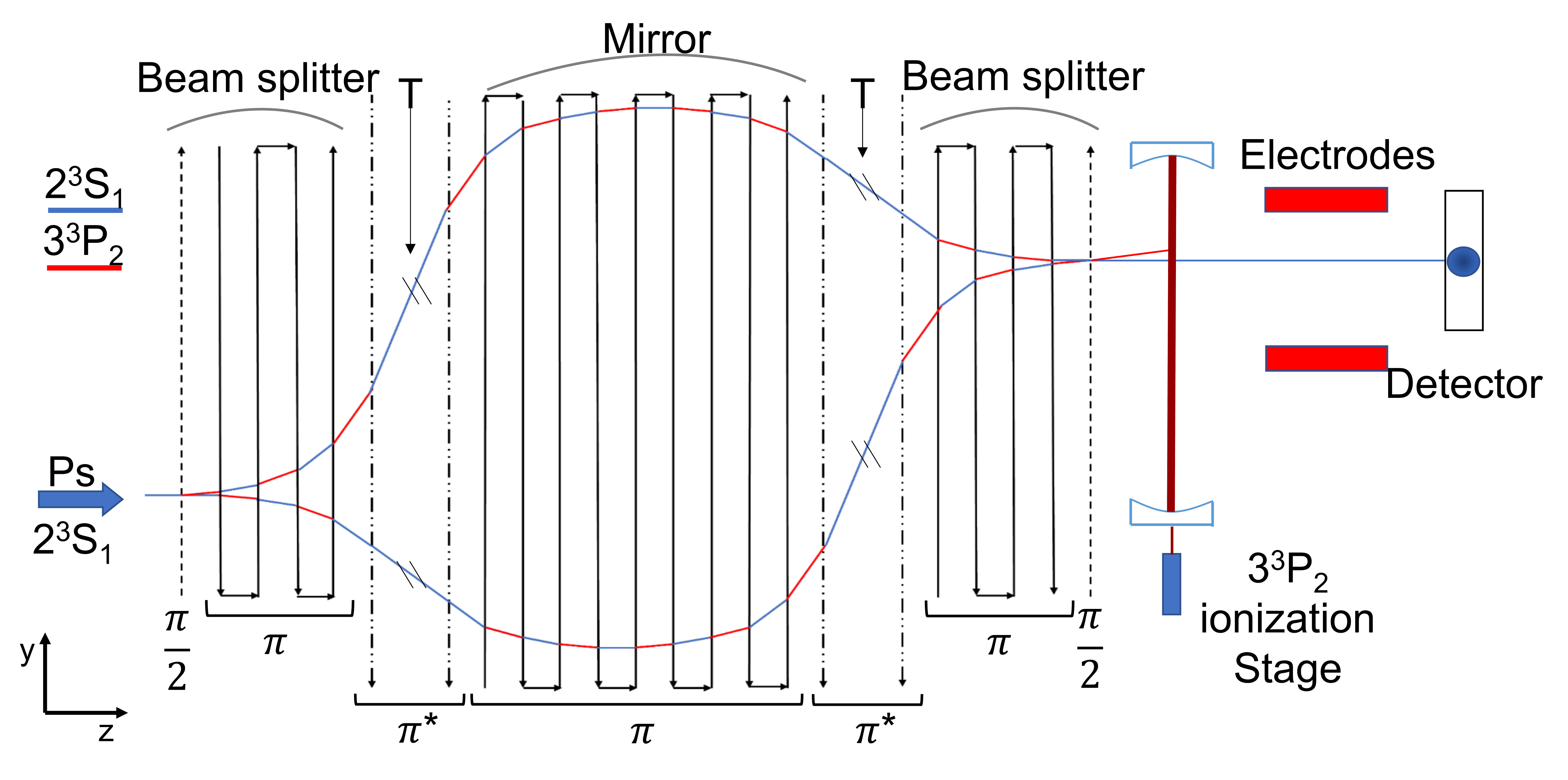}
    \caption{Scheme of the light pulse interferometer with positronium. The Ps beam entering the apparatus propagates according to a laser-driven Mach-Zehnder scheme, in order to acquire a phase induced by the gravitational field. In the scheme, $\pi$/2 pulses act as beam splitters, $\pi$ pulses interchange the $n=2$ states (in blue) with the $n=3$ states in red, while slightly detuned $\pi$* pulses only act to suppress $n=3$ states during the propagation.}
    \label{fig:interferometer}
\end{figure}

We will now discuss the proposed interferometric configuration.
To obtain a 10 $\hbar k$ momentum separation between the wave functions, the device is composed by 23 pulses performing different tasks. These pulses are shown in \fref{fig:interferometer}: after the first group of pulses (a $\pi$/2 followed by four $\pi$ pulses) which acts as a beam splitter, there is a $\pi$* pulse which has the important role of bringing both branches of the interferometer to the $n=2$ state. In fact, the lifetime of the $n=3$ state is too short to allow particles in one of the two arms to propagate for a time equal to $T$ without annihilating.

After the first interrogation, a set of $\pi$ pulses acts as mirrors and precedes two more $\pi$* pulses needed to prepare the second part of the interrogation. At the end of the interferometer, the second beam splitter function is performed again by $\pi$ and $\pi$/2 pulses, used to recombine the wave functions. The working principle of the interferometer has been presented in Ref. \cite{PhysRevLett.124.083604}.
By setting the Rabi frequency of the $\pi$* pulse as $\Omega=\frac{N_{\rm k} \hbar k^2}{m\sqrt{3}}$, where $N_{\rm k}$ is the number of transferred momenta  and $m$ is the Ps mass, it is possible to make it act as a $\pi$ pulse on one of the two arms of the interferometer and as a 2$\pi$ pulse on the other \cite{PhysRevLett.124.083604}. 

For the correct operation of the $\pi$* pulse, it is necessary that the Ps transition probability is zero for a given value of the atom momentum, which can be obtained by a suitable shaping of the laser spatial profile, in the form of a square pulse. This can be implemented by means of top-hat shaping lenses \cite{10.1117/12.907914}; given the intensity of the flat-top laser beam in the form $I(r) = I_0 e^{-2(r/w)^l}$, where $w$ is the beam waist and $l>2$, the simulations have shown that the value $l=6$ would be adequate for an efficient $\pi$* pulse \cite{10.1117/12.930284,Ngcobo:13}. $\pi$ and $\pi$/2 pulses have the same wavelength of 1312 nm, while $\pi$* pulses are slightly detuned from the transition to work at the maximum transition probability for the $n=3$ states.

After crossing the interferometer, the weakly bound $n=3$ state is laser-ionized \cite{PhysRevA.94.012507} by means of an Erbium fiber laser at 1560 nm in order to allow the measurement of the number of $n=2$ Ps atoms alone (\fref{fig:setup}). The still charged particles (electrons and positrons) are swept away by means of a moderate electric field. This choice is motivated by the fact that, without the removal of the $n=3$ level, in order to distinguish between the $n=2$ and $n=3$ populations, one would need to drastically increase the spatial distance between them requiring the addition of a considerable number of pulses (and linear space) to the interferometer. At least eight $\pi$ pulses and one meter of propagation would be required downstream, admitting a maximum angular divergence of the order of few tens of microradians. Given the speeds involved, this constraint would require an additional 3 meters of upstream collimation resulting in a substantial Ps beam population loss due to annihilation. Given the interferometer's tendency to scatter atoms that have not interacted properly with the laser pulses, the signal-to-noise ratio at the detector can be greatly improved by affixing a physical mask that selects the area with the highest signal concentration.

\begin{figure}[ht]
    \centering
    \includegraphics[width=1\textwidth]{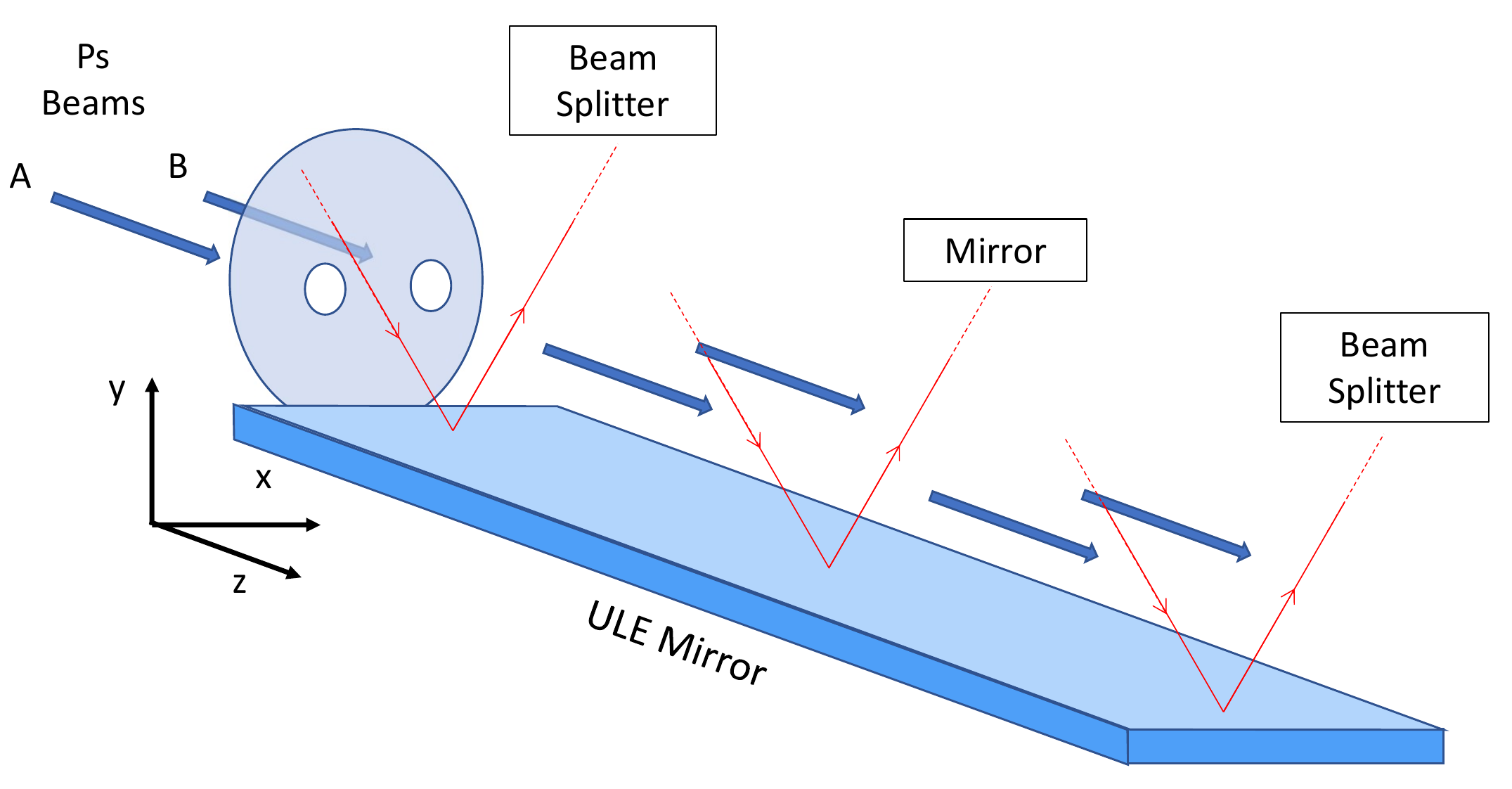}
    \caption{Scheme of the differential measurement strategy on the subdivided Ps beam. The laser radiation hits the two beams A and B with opposite momentum to reject the phase noise. For simplicity, only three pulses are shown here, but the strategy is designed for all the 23 pulses of the interferometer.}
    \label{fig:DoubleInt}
\end{figure}

Since the $n=3$ atoms do not need to be counted, detector spatial resolution is not required. There are several possibilities for this type of detection. Some of these are scintillators, microchannel plate detectors, which are still sensitive to the atomic velocities involved \cite{Vinelli_2020}, or germanium detectors \cite{doi:10.1063/1.5135842,doi:10.1063/1.123815}.
Such a structured interferometer has no control over the laser phase noise and would require an accuracy on the positioning of the optical components that could exceed the experimental possibilities. The same problem arises in the reference literature of gravitation with Ps \cite{OBERTHALER2002129,Mariazzi2020} where sub-nm accuracy is required in gratings placement. It is proposed to solve the issue by extracting the phase signal with a differential method \cite{Rosi2017,PhysRevA.89.023607}: in order to minimize the effect of the laser phase noise, a double measurement can be performed by splitting the Ps beam in two and having the laser pulses hitting the beams from opposite directions, as is shown in \fref{fig:DoubleInt}.

This configuration works effectively as a couple of Positronium interferometers, each making use of two identical beam; two independent $n=2$ population set can therefore be measured during the same data taking, so that the phase $\Phi$ can be extracted from the Lissajous plot of the two combined populations, similar to what was done for atomic gravitation in Ref. \cite{Rosi2017}.

The advantage of this strategy is the rejection of the common mode noise that does not depend on the laser wave-vector and the doubling of the signal which becomes proportional to $2k_{\rm eff} g T^2$ \cite{Foster:02}.
In our proposed configuration the reflection mirror for the double measurement covers the whole length of the interferometer and is suspended by anchors designed to minimize deformations.
Given its high spatial extension, the interferometer could be subject to thermal drifts and mechanical stresses, especially on the reflection mirror. It is possible to significantly reduce these effects by making the mirror with ultra-low expansion (ULE) material.
Any other interferometer phase shifts that depend on $k$ but not on $T$ (e.g., possible deformation of the mirror in the transverse direction) can be reduced by following the described sequence with two different Ps velocities, e.g., $6\times10^6$ m/s and $1.2\times10^7$ m/s, obtained by changing the acceleration potentials of the electrodes.
When the interferometer is run with twice the Ps speed, the interrogation time scales from $T$ to $T/2$ and by measuring the difference between the two cases of slow and fast beam, one obtains a 1.5 final gain on the signal:
\begin{eqnarray}
\eqalign{ \Delta\Phi=\Delta\Phi_{\rm Slow}-\Delta\Phi_{\rm Fast}\\=2k_{\rm eff} gT^2+\delta\Phi(k_{\rm eff})-[2k_{\rm eff}g\left(\frac{T}{2}\right)^2+\delta\Phi(k_{\rm eff})]  =\frac{3}{2} k_{\rm eff} gT^2}
\end{eqnarray}

Once the phase shifts of both measurements have been obtained by analyzing the two differential signals, the final value of the phase difference is obtained by simple subtraction.

\section{Numerical simulation of the LMT interferometer}\label{Sec:Sim}
\begin{figure}[ht]
    \centering
    \includegraphics[width=1\textwidth]{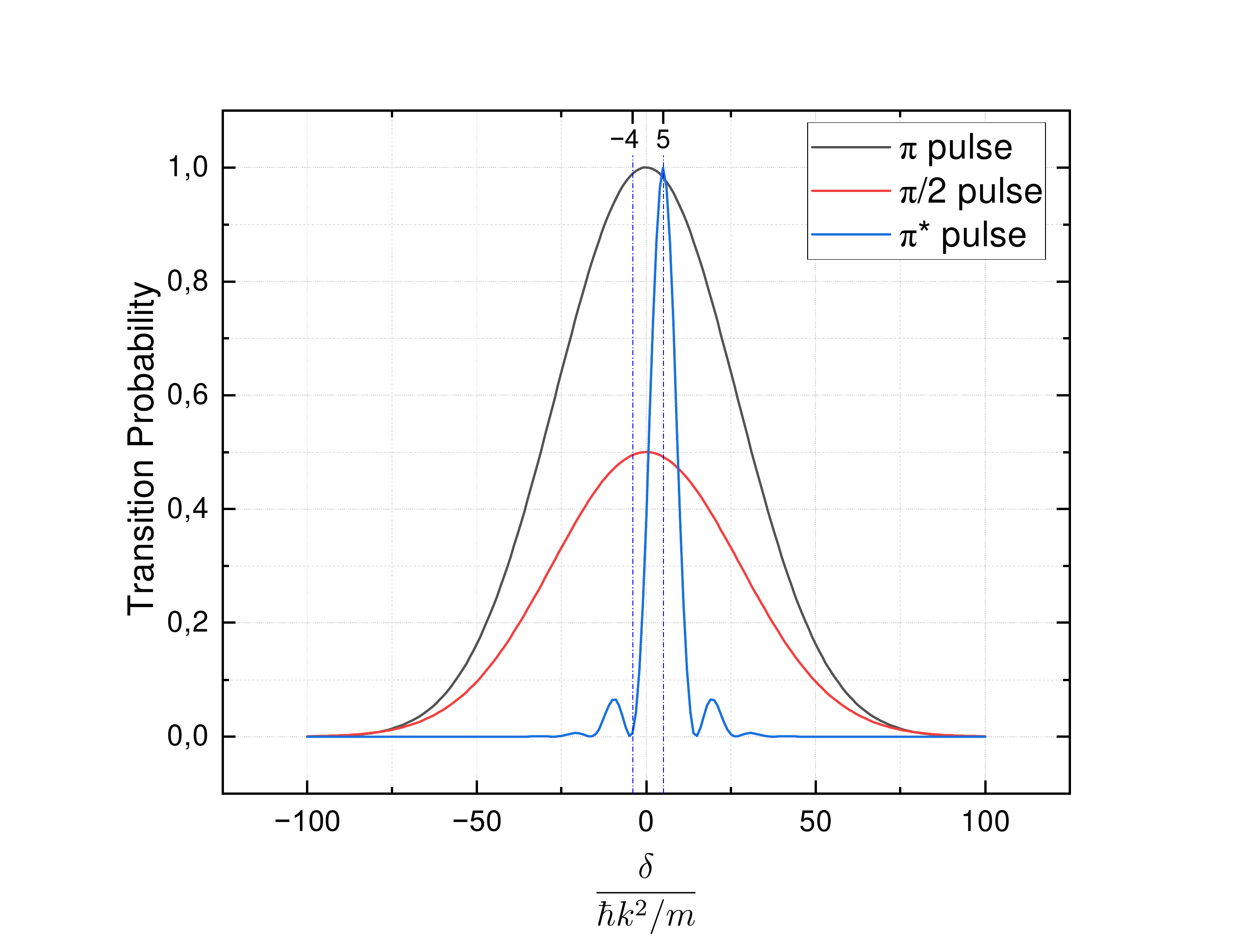}
    \caption{Transition probability between the $n=2$ and $n=3$ states as a function of the ratio between the detuning and the recoil frequency. The $\pi$* pulse detuning is also shown: it acts as a 2$\pi$ pulse for the interferometric wave function which absorbed 4 photons with negative momentum, giving zero transition probability, and as a $\pi$ pulse for the wave function which absorbed 5 photons with positive momentum.}
    \label{fig:PulseEff}
\end{figure}

In this section we briefly summarize the methods used to simulate the interferometer. A more detailed description is given in the appendix.
The core part of the simulation concerns the interaction between the laser pulses and the atomic wave functions defined by
\begin{eqnarray}
\ket{\Psi}=\sum_q c_{\textrm{g},2q}\rme^{-\rmi\omega_{\textrm{g},2q}t}\ket{g,2q\hbar k}+c_{\textrm{e},2q+1}\rme^{-\rmi\omega_{\textrm{e},2q+1}t}\ket{e,(2q+1)\hbar k}
\end{eqnarray}
The expression $\ket{\alpha,p}$ refers to a generic state, where $\alpha$ indicates the energy level (ground, g, or excited, e, which correspond to 2$^3$S$_1$ and 3$^3$P$_2$ respectively) and $p$ is the momentum. $\omega$ is the angular frequency which defines the internal atomic energy $\hbar\omega_{\alpha,p}=\hbar\omega _\alpha +\frac{p^2}{2m}$, and $q$ is the momentum index ($q\in \mathbb{Z}$). Since in the center-of-mass reference frame the initial Ps state is $\ket{g,0}$, $p$ is even for ground states and odd for excited states. In general, the summation over $q$ is infinite, but only the states having $q$ between $-(N-1)$ to $N-1$ will be populated, where $N$ is the number of pulses. The amount of momentum states to be considered for the simulations is therefore $2N$.
The system is described by the sum of the atomic unperturbed Hamiltonian $\hat{H}_0=\hbar\omega_{\rm g}\ket{g}\bra{g}+\hbar\omega_\rme\ket{e}\bra{e}+\frac{\hat{p}^2}{2m}$ and the atom-laser interaction Hamiltonian $\hat{H}^{\uparrow\downarrow}_L=-\frac{E_0(t)}{2}(\rme^{\rmi(\pm ky-\omega_{\rm L}t+\phi)}+\rme^{-\rmi(\pm ky-\omega_{\rm L} t+\phi)})D$, where $D$ is the transition dipole moment, $E_0$ is the amplitude of the laser electric field, $\omega_{\rm L}$ is the frequency of the laser and the $\pm$ sign accounts for the laser propagation direction. It has been assumed that the laser propagates along the $y$ axis. The result is a two-level Schrödinger equation with a variable Rabi frequency because of the Gaussian shape of the laser beam
\begin{eqnarray}
\begin{cases}
    \dot{c}_{\rm g}=\frac{-\rmi\Omega(t)}{2}c_\rme\rme^{\rmi(\delta t-\phi)}\\
    \dot{c}_\rme=\frac{-\rmi\Omega(t)}{2}c_{\rm g}\rme^{-\rmi(\delta t-\phi)},
\end{cases}
    \label{eqn:system}
\end{eqnarray}
where $c_{\rm g}$ and $c_\rme$ are the relative probability amplitudes, $\delta$ is the detuning from the transition frequency and $\phi$ is the laser phase. Given the time dependence of the Rabi frequency, the system has to be integrated numerically (e.g. using a 4-th order Runge-Kutta method).
\Fref{fig:PulseEff} reports the transition probability from one state to another, given by $\abs{c_{\rm g}}^2$ or $\abs{c_\rme}^2$, as a function of $\delta$. Since the Ps Doppler shift has the same effect of varying $\delta$, \fref{fig:PulseEff} also indicates the robustness of the interferometer to Doppler effects: the wider the curves, the more robust the interferometer is. The pulse most sensitive to the Doppler effect is certainly the $\pi$*.

The curves correspond to $\pi$ and $\pi$/2 pulses having waist in the $x$ direction (perpendicular to both Ps and laser propagation) equal to 1.4 mm, and 0.16 mm in $z$ (parallel to Ps propagation), and to $\pi$* pulses with waist equal to 1.8 mm and 1.13 mm in $x$ and $z$ respectively. Increasing the power (and decreasing the waist) widens the curves. The selected optical pulse powers are about 25 W, 6.5 W and 4 W for the $\pi$, $\pi$/2 and $\pi$* pulses respectively and are achievable with coherent combination techniques \cite{Chiow:12,Seise:10}. 
We define the interferometer efficiency as the product between the $\pi$/2 pulse efficiency and the geometric mean of the mirror efficiencies of the upper and lower branches: $\eta=\eta_{\pi/2}\sqrt{\eta_{\rm top}\eta_{\rm bot}}$. Given that the pulse transition probability after an atom-pulse interaction, P$_{\rm t}$, is the square modulus of the probability amplitude of the target atomic state (for instance, e) normalized by the probability amplitude of the starting atomic state (for instance, g), we can define the efficiency of the upper and lower branch as the product of the transition probabilities of all the $\pi$ and $\pi$* pulses in the branch: $\eta_{\rm top}=\prod_{\rm top}P_{\rm t}^\pi P_{\rm t}^{\pi^*}$, $\eta_{\rm bot}=\prod_{\rm bot}P_{\rm t}^\pi P_{\rm t}^{\pi^*}$. Instead, the  efficiency of the splitters is defined by $\eta_{\pi/2}=4\sqrt{(1-P_{\rm t,1}^{\pi/2} ) P_{\rm t,1}^{\pi/2} (1-P_{\rm t,2}^{\pi/2}) P_{\rm t,2}^{\pi/2}}$, where $P_{\rm t,1}^{\pi/2}$ and $P_{\rm t,2}^{\pi/2}$ are the transition probabilities of the two $\pi$/2 pulses.  $\eta_{\rm top}$,  $\eta_{\rm bot}$ and $\eta_{\pi/2}$  represent the probability of an atom to interact correctly with the pulses in the upper and lower branches and in splitter stages respectively. Note that all probabilities $P_{\rm t}$ vary along the interferometer due to the Doppler effect resulting from momentum transfer from the pulses to the Ps. 

\begin{figure}[ht]
    \centering
    \includegraphics[width=1\textwidth]{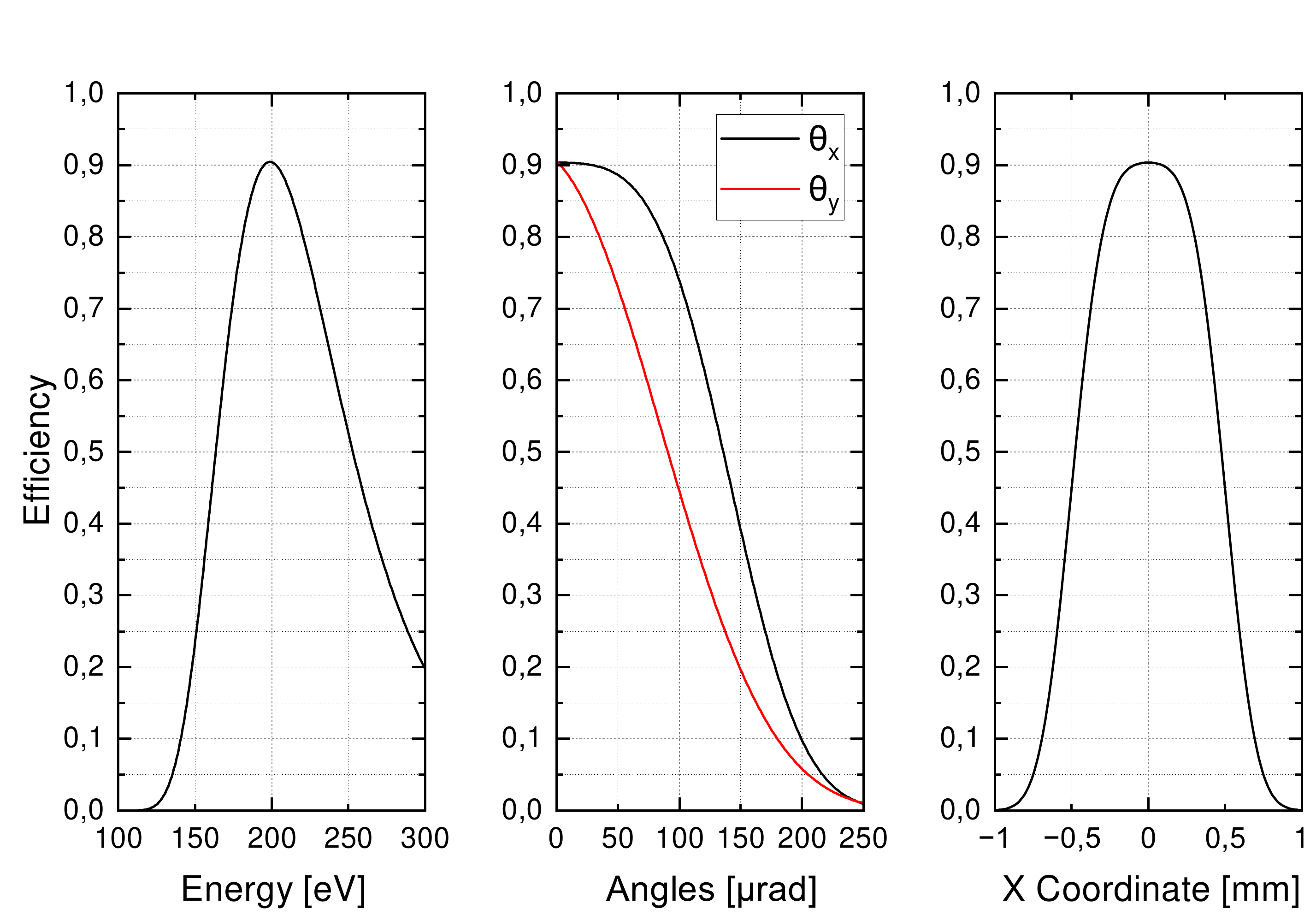}
    \caption{Interferometer efficiency as a function of Ps energy, entrance angles and position.}
    \label{fig:EffVsParam}
\end{figure}

In order to have an estimate of the admissible divergence, the efficiency as a function of Ps entrance angles and coordinates was studied, and the results are presented in \fref{fig:EffVsParam}: the efficiency is almost 0 at about 250 $\mu$rad while it is still high for an atom entering the interferometer at 0.25 mm from the centre of the entrance hole.

The dependence of efficiency on Ps energy begins to decrease significantly at about 25 eV from the design value (200 eV), and since Ps is expected to have an energy spread of the order of 10 eV \cite{Sacerdoti,Michishio2019,Meshkov,TrezziAegisPhD}, which corresponds to a decrease in efficiency of less than 3\%, it is not critical in this experiment. Based on these results and by trying different values, an optimal size of the mask to be placed in front of the detector so as to select input angles less than 100 $\mu$rad has been found. 

\section{Results}\label{Sec:Res}
In the proposed configuration the mask in front of the detector plays the important role of selecting atomic trajectories that typically tend to have a favourable interaction with the whole interferometer. Since Ps atoms failing the correct propagation have a lower probability of being accepted, the mask allows to clean up the accepted statistics in terms of fringe visibility (interferometric contrast), which can be set almost independently of the beam distribution.

\begin{figure}[ht]
    \centering
    \includegraphics[width=1\textwidth]{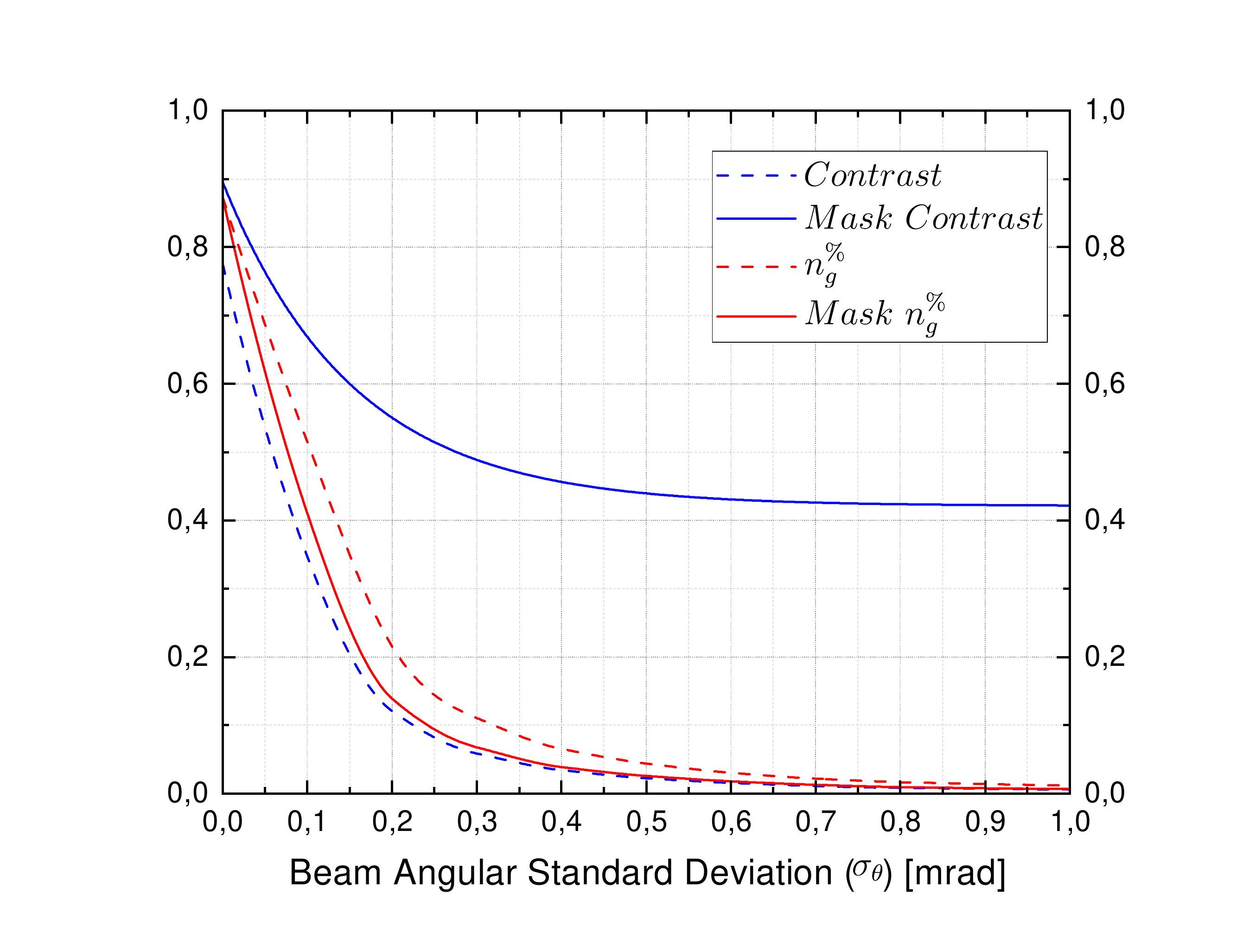}
    \caption{Contrast (blue) and percentage of atoms that have correctly crossed the interferometer (red) vs Ps beam angular standard deviation. The solid lines represent the statistics in a 1.5x1 mm$^2$ acceptance mask while the dashed ones correspond to the case of the whole screen (no mask). The contrast in the mask remains approximately constant and about 0.4 while the contrast of the whole detector tends to 0. The small difference between the percentage of “good” atoms (atoms that correctly crossed the interferometer) outside and inside the mask indicates the suitability of mask sizing.}
    \label{fig:Contrast}
\end{figure}

If $n_{\rm g}$ is the number of atoms that correctly cross the interferometer and $n_{\rm n}$ is the number of atoms that fail the interferometer, we can define the percentage $n_{\rm g}^\%=\frac{n_{\rm g}}{n_{\rm g}+n_{\rm n}}\%$.
In \fref{fig:Contrast} the contrast (blue curves) and $n_{\rm g}^\%$ (red curves) as a function of the Ps beam angular standard deviation, $\sigma_\theta$, in presence (solid lines) and absence (dashed lines) of the mask are shown. Each point of the curves represents a run of the Monte Carlo simulation with different values of $\sigma_\theta$. After an initial decrease, the contrast with the mask stays approximately constant and equal to 0.4 while it tends to 0 without the mask.
The difference between the percentage of signal with and without the mask is low, implying that the mask is well sized since it discards only a small part of the signal but a large amount of noise. The effect of Ps decay is taken into account in the simulation, as can be appreciated by the difference between the initial values of the contrasts for a null $\sigma_\theta$ (perfectly collimated beam). In this case there is still some noise generated by decayed atoms reaching the detector at nonzero angles due to the decay recoil. 

The minimum detectable acceleration value is defined by the sensitivity $\Delta g=\frac{1}{C\sqrt{N_{\rm Ps}}k_{\rm eff}T^2}$ \cite{OBERTHALER2002129} where $N_{\rm Ps}$ is the number of detected atoms and $C$ is the contrast. Note that the mask reduces $N_{\rm Ps}$ but this effect is counteracted by the increase in $C$.
It is reasonable to assume that the beam can fit a Gaussian distribution with $\sigma_\theta$ ranging from 1 to 10 mrad on the angles and a uniform distribution extending for 0.25 mm in diameter on the entrance coordinates. Given the differential measurement with two beams carrying the same number of atoms and taking into account the error propagation, the sensitivity to the gravitational acceleration $g$ becomes
\begin{eqnarray}
\Delta g=\frac{\sqrt{2}}{2C\sqrt{N_{\rm Ps}}k_{\rm eff}T^2}
\end{eqnarray}

 The acquisition time is defined by
 \begin{eqnarray}
t=\frac{N_{\rm Ps}}{\Phi_{\rm Ps}}\frac{1}{\epsilon}=\left(\frac{\sqrt{2}}{20kT^2 \Delta g C}\right) ^2\frac{1}{\epsilon\Phi_{\rm Ps}}
\end{eqnarray}
where $\Phi_{\rm Ps}$ is the Ps flux and $\epsilon$ is the ratio between the number of atoms entering the mask and the total number of atoms entering the interferometer. $\epsilon$ considers the annihilation and the probability that an atom ends up in the mask: $\epsilon=\epsilon_{\rm ann}\epsilon_{\rm mask}$.

Table \ref{SigEpC} shows $\epsilon$ and contrast for different atomic beam divergences. For sigma greater than about 0.5 mrad, the mask begins to cut the beam at the interferometer exit while keeping constant the contrast and reducing $\epsilon$.
Note that the proposed detection method allows the interferometer to perform a "self-cleaning action" by removing from the mask the atoms that have lost the gravitational information. This effect is expressed by the behaviour of $\epsilon_{\rm mask}$: as $\sigma_{\theta}$ increases, $\epsilon_{\rm mask}$ decreases by an amount not solely attributable to the different shape of the atomic beam but also to the scattering of the laser pulses. The advantage of this method is that while $\epsilon$ decreases, the contrast, on which the data acquisition time has a quadratic dependence, remains constant. 
\begin{table}
\caption{\label{SigEpC}$\epsilon$ ratio and interferometric contrast for different values of the beam angular standard deviation. All values have a relative error lower than 5\%.}
\begin{indented}
\item[]\begin{tabular}{@{}lll}
\br
$\sigma_\theta$ [mrad] & $\epsilon$ & C \\
\mr
0.1 & 14\% & 60\% \\
1 & 0.3\% & 40\% \\
3 & 0.03\% & 40\% \\
5 & 0.01\% & 40\% \\
7 & 0.006\% & 40\% \\
10 & 0.003\% & 40\% \\
\br
\end{tabular}
\end{indented}
\end{table}

\begin{figure}[ht]
    \centering
    \includegraphics[width=1\textwidth]{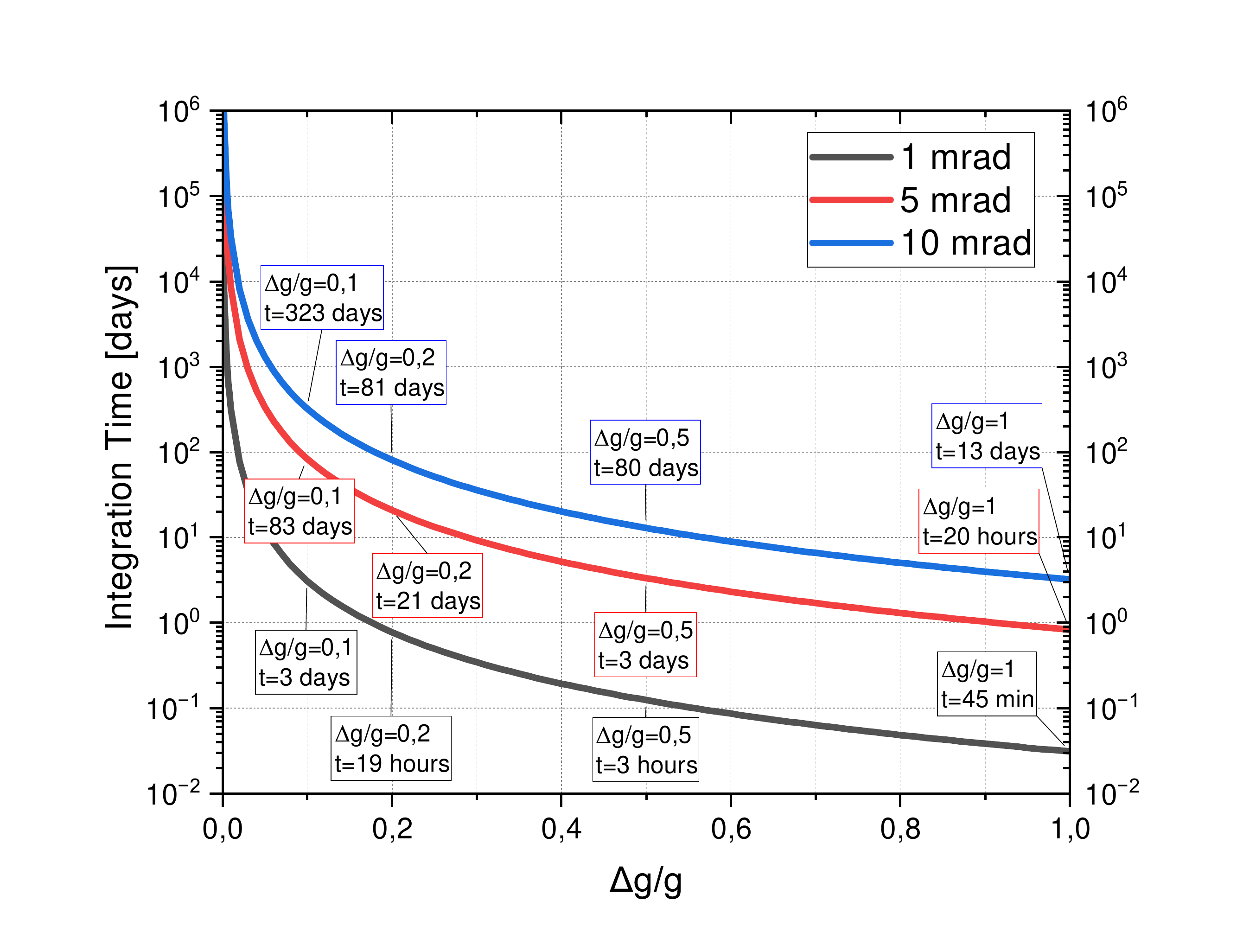}
    \caption{Behaviour of the data acquisition time as a function of the relative error on gravitational acceleration, for three values of the atomic beam angular standard deviation.}
    \label{fig:IntTim3Std}
\end{figure}

With an input flux of 0.5x10$^8$ Ps/s and $\sigma_\theta=10$ mrad, one would obtain $\Delta g/g = 10$\%, which is significant for a fundamental physics test (see section \ref{Sec:PosGrav}), in about 11 months of data acquisition. \Fref{fig:IntTim3Std} shows the signal acquisition time as a function of the desired sensitivity for $\sigma_\theta$ equal to 1, 5 and 10 mrad.

\section{Comparison between Bragg and single photon LMT interferometers}\label{Sec:Com}
\begin{figure}[ht]
    \centering
    \subfigure[]{\includegraphics[width=0.6\textwidth]{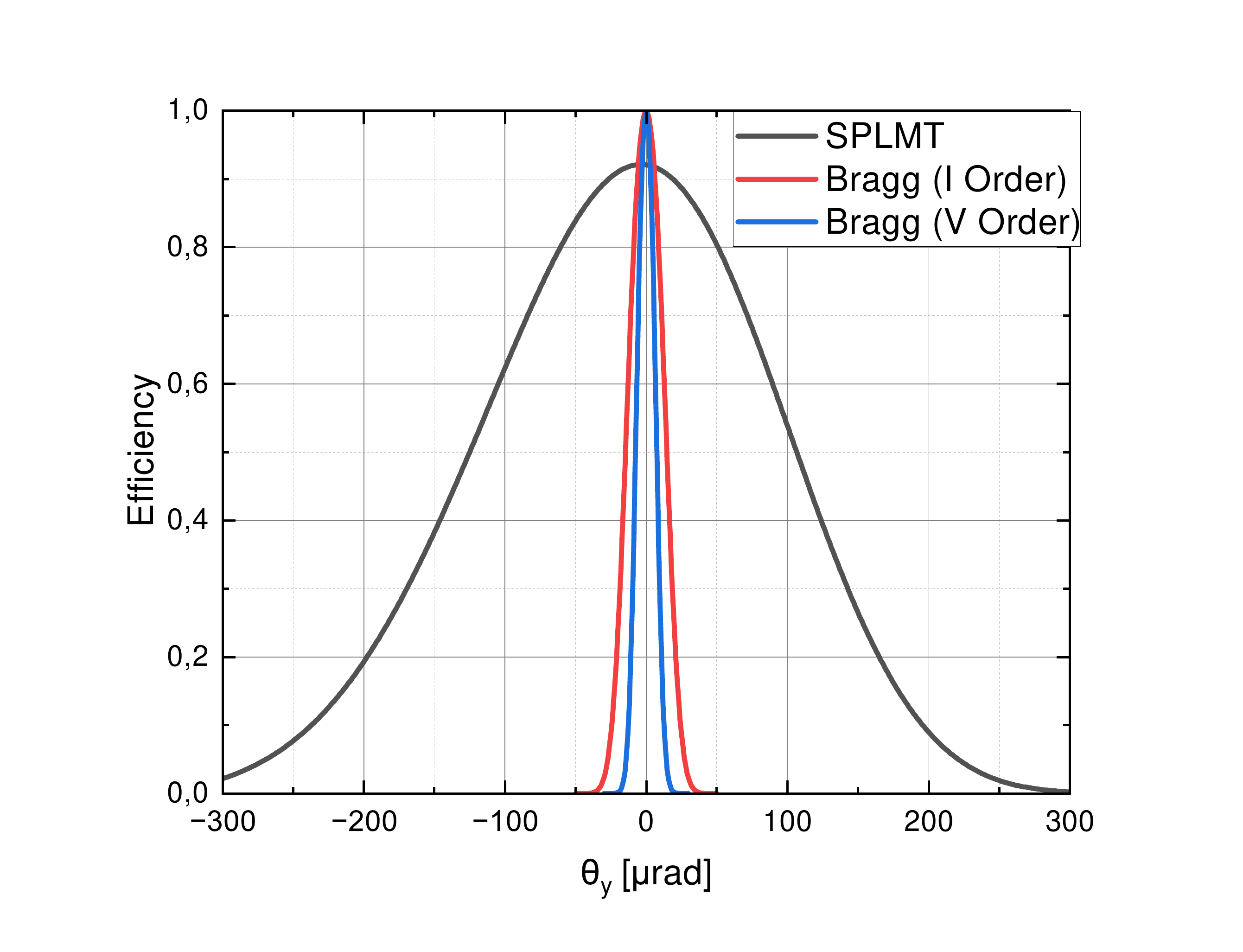}}
    \subfigure[]{\includegraphics[width=0.6\textwidth]{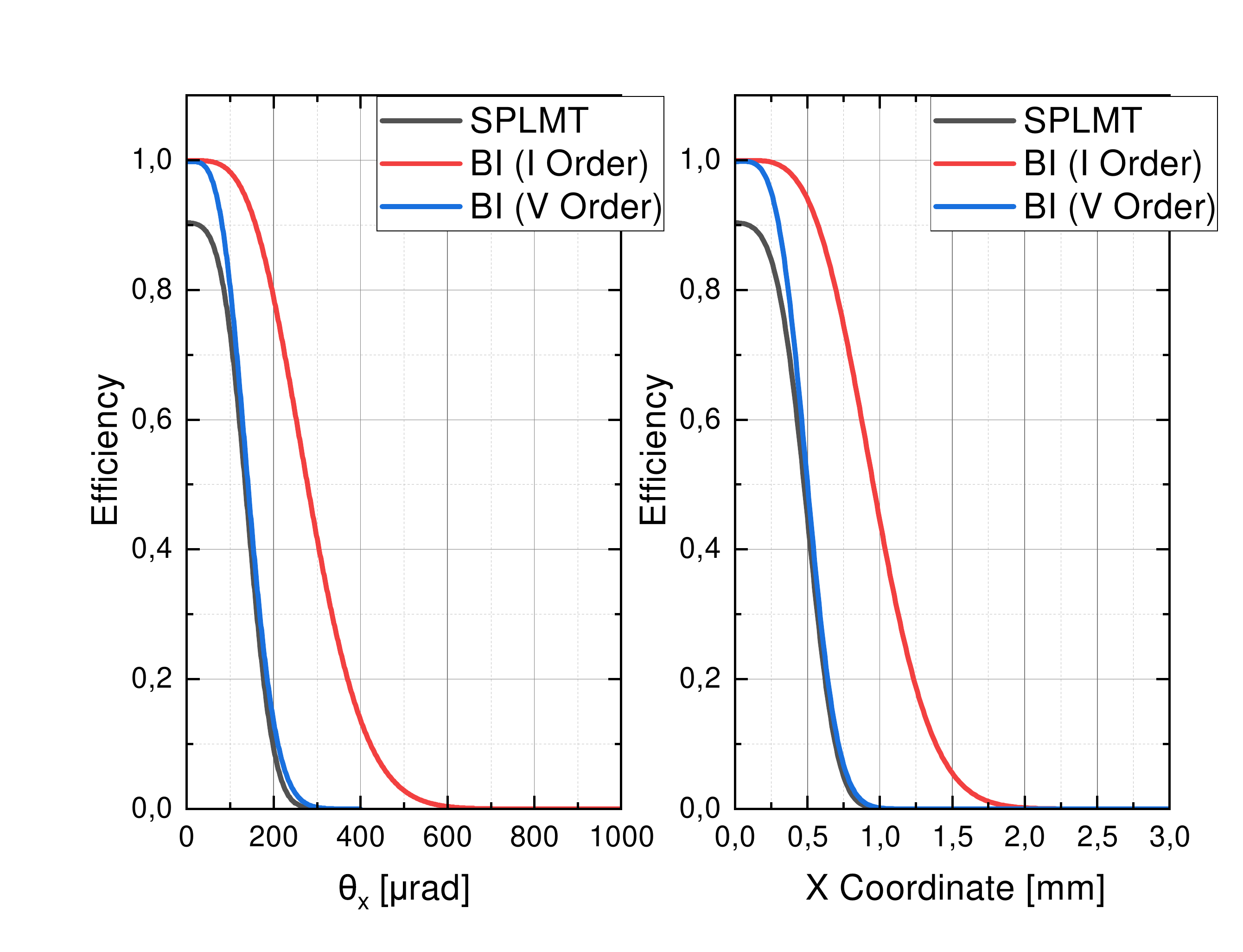}}
    \caption{Efficiency of the interferometers as a function of the entrance angles and position. The curves refer to different interferometric configurations: the Single Photon LMT interferometer discussed in this work (black), and the Bragg interferometers at the first (red) and the fifth (blue) diffraction order. The degree of angular acceptance of Bragg interferometers decreases as the order of diffraction increases. The SPLMT interferometer has the best angular acceptance in the y direction which is the most critical.}
    \label{fig:SPVsBragg}
\end{figure}

In this section we compare the interferometer proposed in this paper with the literature, with reference to the Bragg interferometer (BI) proposed in Ref. \cite{OBERTHALER2002129}. In this regard, we have designed and simulated a similar interferometer: a standard Mach-Zehnder configuration (two $\pi$/2 pulses and one $\pi$ pulse) in Bragg regime at the first and fifth diffraction order, with detuning from the 2S-3P transition of the order of tens of GHz. To reduce the acquisition times, it is necessary to maximize the product $Ck_{\rm eff} T^2$. For the comparison, we have designed a BI with the same $T$ as the single photon LMT (SPLMT). The Ps speed must also be equal because it strongly affects the atomic flux in the focusing stage. This means that the lengths of the interferometers are the same but the BI waist along the z-direction must be much larger, about 7 mm. The constraint on the waist is given by the recoil frequency which sets a minimum value of the interaction time between the atoms and the pulses. The maximum power required by the BIs at the considered diffraction orders ranges from 90 W to 15 kW. These powers with such laser beam sizes pose a major technological challenge, but we will neglect this aspect in this discussion to make an ideal comparison.
The effective momentum is fixed by the design of the interferometers which is 2$\hbar$k for the first BI diffraction order and 10$\hbar$k for the SPLMT and the fifth BI order. 
We have simulated the interferometers to analyze their efficiency as a function of the entrance angles and transverse coordinate, like in \fref{fig:EffVsParam}. The results are shown in \fref{fig:SPVsBragg}. The model used in the BIs simulation is described in Ref. \cite{PhysRevA.77.023609}.
The efficiency of the BIs for $\theta_{\rm y}$=0 is higher because of the smaller number of pulses forming them (3 vs. 23), however, as soon as an imperfectly collimated atomic beam is considered, the Bragg efficiency collapses below that of the SPLMT interferometer due to the Bragg condition becoming more stringent at high atomic velocities. As the diffraction order increases, the angular acceptance decreases, and the efficiency drops to zero for lower values of $\theta_{\rm y}$. Note that this characteristic must also be considered in the alignment of the laser pulses, which must then respect a maximum relative angle with the atomic beam. The efficiency of BI as a function of input parameters related to the $x$ direction is higher for first-order and equal to the SPLMT for fifth-order.

\begin{figure}[ht]
    \centering
    \includegraphics[width=1\textwidth]{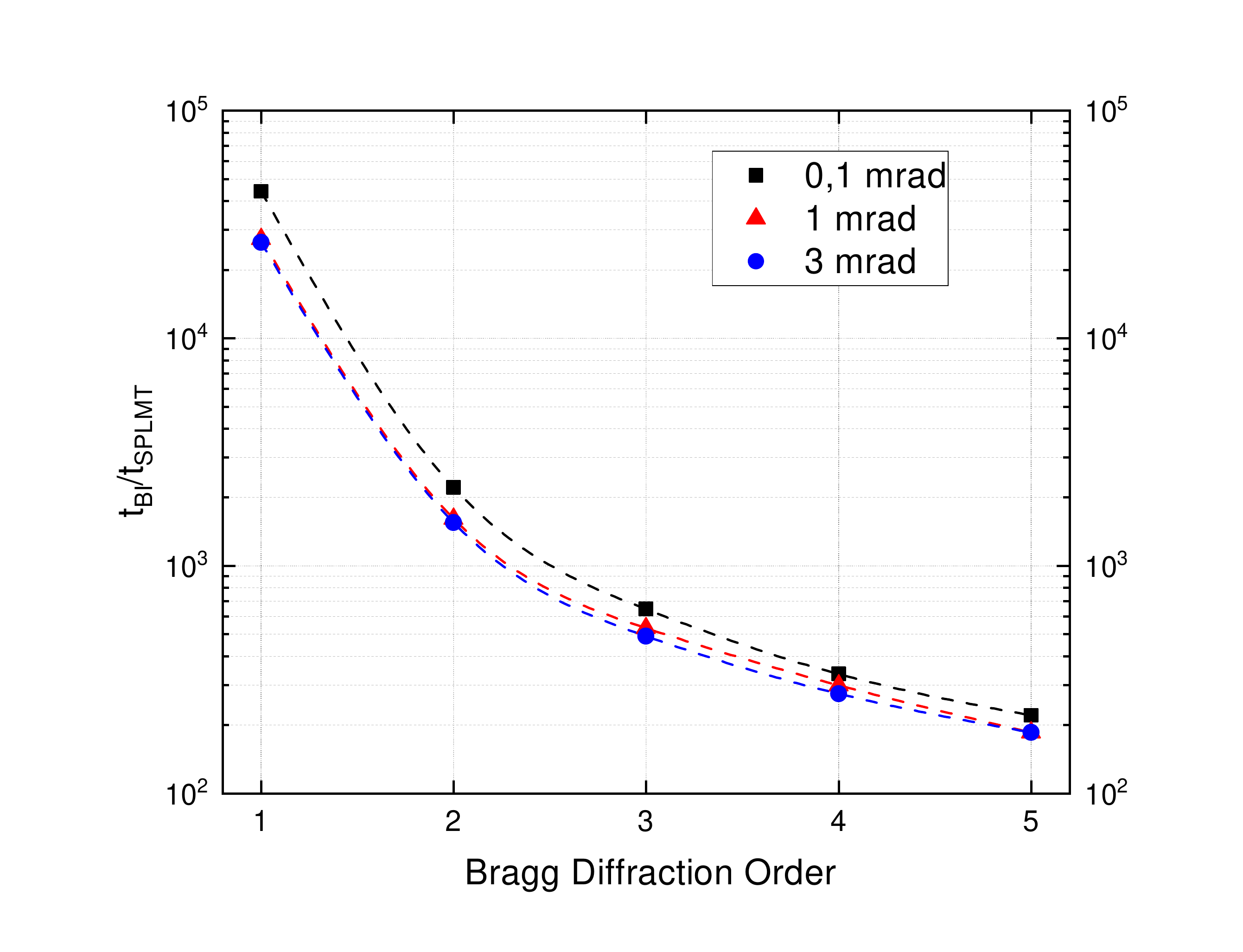}
    \caption{Data acquisition times ratio (Bragg over SPLMT) as a function of the first five Bragg diffraction orders, for three values of beam angular standard deviation. The rate decreases with the diffraction order and  it is much greater than one for every order and $\sigma_\theta$. The relative errors are lower than 5\%.}
    \label{fig:TimeRatio}
\end{figure}

The time of flight through the interferometers is the same but the BIs do not treat $n=3$ states resulting in a different probability of survival to Ps beam decay, $\epsilon_{\rm ann}$.
One of the main limitations of the BI is the impossibility of filtering the interferometric outputs based on the internal state. In fact, the atoms leaving the apparatus are in different momentum states but in the same internal state. This implies that in order to separate the arms and ensure that the signal spots do not overlap on the detector, several metres of downstream propagation could be necessary, resulting in a further loss due to annihilation (i.e. a further reduction of $\epsilon_{\rm ann}$), as already explained in section \ref{Sec:LMTI}. The separation of the spots must be sized according to the desired degree of angular acceptance and it is defined by
\begin{eqnarray}
Z_{sep}=[D_{\rm y}+2Z_{\rm I} tg(\theta_{\rm y}^{\rm max})] \frac{v_{\rm z} m}{ N_{do} \hbar k}
\end{eqnarray}
where $D_y$ is the entrance hole diameter, $Z_{\rm I}$ is the interferometer length, $\theta_{\rm y}^{\rm max}$ is the maximum accepted angle of the atomic beam and $N_{\rm do}$ is the diffraction order. Choosing $D_{\rm y}=0.25$ mm and $\theta_{\rm y}^{\rm max}=125$ $\mu$rad (as for the SPLMT), the propagation needed after the interferometer ranges from 32 m to 6.5 m from the first to the fifth order. As already mentioned, it is necessary to ensure that atoms with an angle greater than $\theta_{\rm y}^{\rm max}$ do not enter the interferometer via upstream collimation. This requires another 2.5 m of propagation. The total distance travelled by an atom exiting the microwave cavity would be 39.5 m and 14 m for the first and fifth diffraction order, respectively. $\epsilon_{\rm ann}$ would be about 0.32\% for the first order and 14\% for the fifth order that should be compared with $\epsilon_{\rm ann}\sim27$\% of the SPLMT. 
In the Bragg configuration no mask is needed but the collimation reduces the atomic flux: $\epsilon_{\rm BI}$=$\epsilon_{\rm ann}\epsilon_{\rm coll}$. The beam is also collimated in the $x$ direction so that the BIs operate at a minimum efficiency of about 30\%: this selects a maximum angle and input coordinate that varies with diffraction order (see \fref{fig:SPVsBragg}, b).
The ratio between the acquisition times of the two types of interferometers is given by
\begin{eqnarray}
\frac{t_{\rm BI}}{t_{\rm SPLMT}}=\left(\frac{k_{\rm SPLMT}^{\rm eff} C_{\rm SPLMT}}{k_{\rm BI}^{\rm eff} C_{\rm BI}}\right)^2\frac{\epsilon_{\rm SPLMT}}{\epsilon_{\rm BI}}
\end{eqnarray}
This ratio is shown in \fref{fig:TimeRatio} as a function of the BI's diffraction order for three angular standard deviation values. As the order increases, the ratio decreases due to the higher effective momentum and the lower separation length at the detection. This decrease is counteracted by the lower angular acceptance which is inversely proportional to the diffraction order (as shown in \fref{fig:SPVsBragg}). The higher value of the ratio at 0.1 mrad is due to the fact that, as $\sigma_\theta$ decreases, the contrast of SPLMT starts to increase earlier and more than that of BI. The difference between the curves at 1 and 3 mrad is not significative beacuse for $\sigma_\theta\geq 1$ mrad the contrasts and $\epsilon_{\rm SPLMT}/\epsilon_{\rm BI}$ are approximately constant. For fifth-order, the time ratio is driven by $C_{\rm BI}/ C_{\rm SPLMT}$ and $\epsilon_{\rm SPLMT}/\epsilon_{\rm BI}$, where $\epsilon_{\rm BI}$ is governed by the decay in the collimation and separation zone. Both ratios favor the SPLMT interferometer emphasizing the main differences between the two types of interferometer: BI is not optimal with high-speed atomic beams because of the minimal interaction time required between light and atom, and the detection mode is not selective on the internal atomic state.  Given the technical difficulties of generating pulses with powers of tens of kW and waist of about 7 mm, we did not find it necessary to analyze and simulate BIs at higher orders. A Bragg interferometer may be feasible and cost-effective with a much slower beam, but a high reduction in speed would lead to a drastic reduction in atomic flux, both within the interferometer and in the focusing zone, resulting in an inconveniently large increase in integration time.
This result shows that using a Single Photon LMT interferometer is more convenient than a Bragg interferometer for positronium inertial sensing.

\section{Conclusions}
In this work we reported the design and the results of the numerical simulation of a single photon LMT interferometer for measuring the gravitational acceleration on a Ps atom. We have defined the characteristics of the interferometer and the laser pulses that compose it and we have studied the energy, angular and position distributions of the atomic beam as Monte Carlo parameters. After having highlighted the criticality of the angular distributions, we proposed a detection based on the ionization and removal of one of the two interferometer arms to solve the problem of the spots overlapping on the detector, and on the noise filtering through a physical mask that selects the fraction of the atomic beam entering the interferometer with an angle smaller than 100 µrad. With this type of detection a good deal of noise can be filtered out, thereby reducing acquisition times. 
It has been shown that to counteract the effects of laser phase noise, it is possible to use a double collinear interferometer which also has the advantage of doubling the signal. Given its high spatial extension, the interferometer could be subject to thermal drifts and mechanical stresses, especially on the reflection mirror. It is possible to significantly reduce these effects by building the mirror with ULE material. To counteract these and other sources of noise that do not depend on the wave vector of the laser pulses, it is possible to carry out the measurement with two different atomic beam velocities. The numerical simulation showed that it would take less than one year of signal acquisition to obtain a relative error $\Delta$g/g=10\% with a Ps beam angular standard deviation equal to 10 mrad. Although it is a fairly long period of time, the interferometer is designed to be as robust as possible to the noise sources usually present in atomic light interferometers and its realization is mainly a technical matter, the main challenge of which stems from the complexity of the apparatus. Furthermore, it cannot be excluded that improvements in the focusing apparatus for the atomic beam will allow in the future to obtain a $\sigma_\theta$ of the order of hundreds of microradians, greatly facilitating the implementation of this experiment.
The work presented here should be compared to the literature, with particular reference to \cite{OBERTHALER2002129}. For this reason, we numerically simulated a BI and compared it with the one proposed here. The results showed that the Single Photon LMT interferometer, coupled with a noise filtering selective method, is more efficient for poorly collimated fast beams, and it requires shorter acquisition times to complete the experiment.
The analysis reported in this work shows that the construction of an interferometer to measure the gravitational effect on antimatter is a complex technical challenge, but the goal is achievable. The efforts spent to set up an apparatus such as the one proposed here would be amply rewarded by the significant scientific achievements that would derive from the first measurement of the gravitational effect on positronium.

\ack
G. Vinelli acknowledges the project CNR-FOE-LENS-2021 for support.

\section*{Data availability}
The data that support the findings of this study are available upon reasonable request from the authors.

\appendix
\section*{Appendix}
\begin{figure}[ht]
    \centering
    \includegraphics[width=1\textwidth]{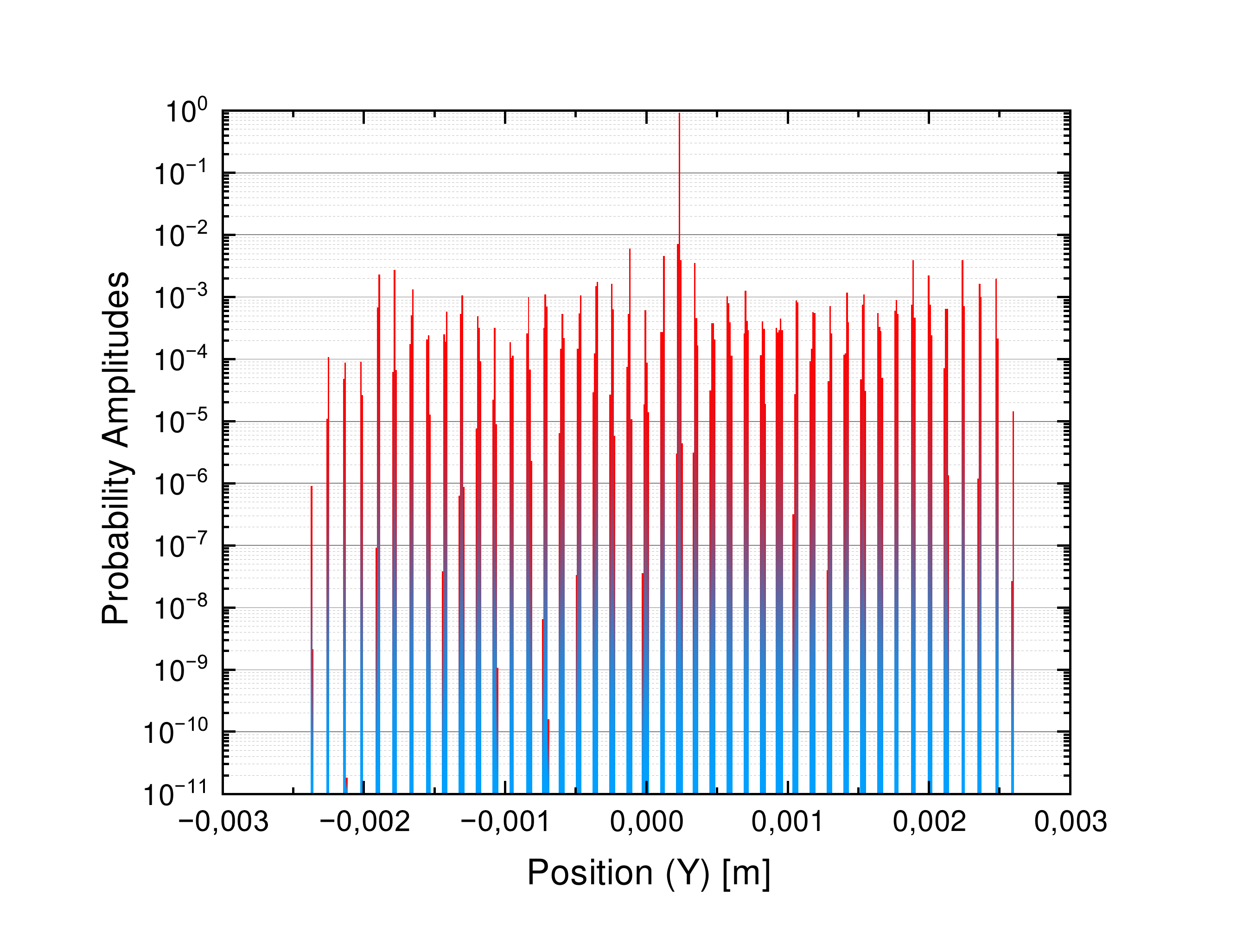}  
    \caption{Probability amplitudes of the states corresponding to each possible trajectory for an atom with null entrance angles. The abscissa axis indicates the arrival position ($y$ coordinate) at the detector of the Ps atoms having different trajectories. The highest bar represents the states that carry the gravitational information and the corresponding amplitude is about 0.92. The other bars represent the noise and the sum of their amplitudes covers the remaining probability, which is equal to 0.08.}
    \label{fig:ProbAmps}
\end{figure}

\setcounter{section}{1}
An atom that crosses the interferometer can be represented by a wave function populating two different energy and momentum states after an interaction with a pulse. The two states separate spatially along the propagation thus defining two distinct trajectories. By applying this process to the entire interferometer, one obtains that the maximum number of trajectories is equal to 2$^N$, which is about 8 million for the interferometer proposed in this work. The effective number is reduced if we consider that two states recombine when they spatially overlap at a laser pulse. From now on we will call the wave function path indicated in \fref{fig:interferometer} as "main pattern" and all other paths as "parasitic pattern" or "parasitic trajectories". The main one corresponds to a proper atom-pulses interaction pattern, and is the most probable if the interferometer is well sized. On the other hand, the parasitic pattern represents all other trajectories arising from losses in the single interactions. The gravitational information is contained only in the main pattern while the parasitic one contributes to the noise. Given the high number of parasitic trajectories, it is not possible to define a priori how many and which of these interfere with each other or with the main pattern, changing its phase and therefore the gravitational information. This issue was studied by a full quantum-mechanical simulation of the interferometer, which independently traces each trajectory and sums up the probability amplitudes in case of spatial superposition inside a pulse (which is in the present case the intersection between the trajectories \cite{NOTA}).
We can generally describe the problem in quantum mechanical terms as
\begin{eqnarray}
\ket{\Psi_n}=\hat{O}_n\ket{\Psi_0},
\label{eqn:PsiImpl}
\end{eqnarray}
where $\Psi_n$ and $\Psi_0$ are the wave functions at the beginning of the interferometer and after the $n$-th pulse, where $n$ ranges from 1 to $N$. $\hat{O}_n$ is the operator expressing the action of the pulses. Starting with Ps in the ground state with zero momentum (with respect to its center-of-mass) we have $\ket{\Psi_0}=\ket{g,0}$, while for $\ket{\Psi_n}$:
\begin{eqnarray}
\ket{\Psi_n}=\sum_{q=-Q_n}^{Q_n}\left( c_{n,\textrm{g},q}(t)\ket{g,2q\hbar k}+c_{n,\textrm{e},2q+1}(t)\ket{e,(2q+1)\hbar k}\right)
\label{eqn:PsiExpl}
\end{eqnarray}
In \eref{eqn:PsiExpl}, $c_{n,\alpha,p}$ is the probability amplitude of the $\alpha$ internal state with momentum $p$ after the $n$-th pulse and $Q_n=\frac{n-1}{2}$ if n is odd or $Q_n=\frac{n}{2}$ if n is even. Note that each state contains a high number of trajectories: $\ket{\alpha,p}=\sum_{j=1}^Jc_{\alpha,p,j}\ket{\alpha,p,j}$, where $j$ is the trajectory index and $J$ is the total number of trajectories which depends on $p$ (e.g., the states in which more trajectories converge are those with $p$ close to 0). Putting it all together we get
\begin{eqnarray}
\fl \ket{\Psi_n}=\sum_{q=-Q_n}^{Q_n}\left( \sum_{j=1}^{J(2q)}c_{n,\textrm{g},2q,j}(t)\ket{g,2q\hbar k,j}+\sum_{j=1}^{J(2q+1)}c_{n,\textrm{e},2q+1,j}(t)\ket{e,(2q+1)\hbar k}\right).
\end{eqnarray}
The interferometric operator can be expressed as the product of the pulse operator $\hat{L}$ times the propagation operator $\hat{P}$ which expresses the free propagation of the wave function after the $n$-th pulse 
\begin{eqnarray}
    \hat{O}_n=\prod_{i=1}^n\hat{L}_i\hat{P}_i
\end{eqnarray}
$\hat{L}$ is a $2N\times2N$ matrix having all the elements of the main diagonal and some elements of the first supradiagonal and subdiagonal different from 0. These elements are computed by solving system \eref{eqn:system} for two coupled states with  $\delta=\omega_{\rm L}-(\omega_{\rm g}-\omega_\rme)+\frac{(2q\hbar k)^2}{2m}-\frac{((2q\pm 1)\hbar k)^2}{2m}$, where the $\pm$ sign depends on the pulse direction. For example, considering a pulse propagating in the negative $y$ direction, the states $\ket{g,2q\hbar k}$ couple with the states $\ket{e,(2q-1)\hbar k}$, thus, just considering $q$ = 0, the two coupled states are $\ket{g,0}$ and $\ket{e,-\hbar k}$. Since system \eref{eqn:system} has no analytical solution, we cannot explicitly write the elements of the operator $\hat{L}$ but we can substitute it with the probability amplitudes of the states of the system that would result from applying the operator by solving \eref{eqn:system} numerically. Collecting these amplitudes into a matrix $\Tilde{L}$, for a pulse with a negative propagation direction and assuming that $N$ is even, the matrix $\Tilde{L}^\downarrow$ takes the form

\begin{eqnarray}
\fl
\begin{bmatrix}
c_{\rm e,-(N-1)}&c_{\rm g,-(N-2)\rightarrow e,-(N-1)}&\dots&0&0\\
c_{\rm e,-(N-1)\rightarrow g,-(N-2)}&c_{\rm g,-(N-2)}&\dots&0&0\\
\vdots&\vdots&\ddots&\vdots&\vdots\\
0&0&\dots&c_{\rm e,N-1}&c_{\rm g,N\rightarrow e,N-1}\\
0&0&\dots&c_{\rm e,N-1\rightarrow g,N}&c_{\rm g,N}
\end{bmatrix},
\end{eqnarray}

where, for instance, $\abs{c_{\rm g,-(N-2)\rightarrow e,-(N-1)}}^2$ is the probability to have a transition from a ground state with $-(N-2)\hbar k$ to an excited state with $-(N-1)\hbar k$ after the pulse. $\abs{c_{\rm e,-(N-1)}}^2$ is the probability to find the atom in the same $\ket{e,-(N-1)\hbar k}$ state before and after the pulse. For a pulse with a negative propagation direction, the matrix elements $[\Tilde{L}_{j,j};\Tilde{L}_{j+1,j}]$ and $[\Tilde{L}_{j,j+1};\Tilde{L}_{j+1,j+1}]$ are solution of \eref{eqn:system} with $[c_{\rme,2q};c_{\textrm{g},2q}]=[0;1]$ and $[c_{\rme,2q-1};c_{\textrm{g},2q-1}]=[1;0]$ as initial condition for the Runge-Kutta integrator.
The propagation matrix is a $2N\times2N$ diagonal matrix whose elements are $e^{-\rmi H_0 t/\hbar}$ (free propagation). 
By computing \eref{eqn:PsiImpl} with a matrix product, it would be impossible to distinguish the probability amplitudes of different trajectories belonging to the same momentum state since they would be mixed (\textit{total mixing}). In this condition, the assumption for which only overlapping trajectories sum up would not be respected. It is therefore necessary to set up a symbolic calculation that uses the propagation operator to distinguish the trajectories and sum up the probability amplitudes with a common propagator only (\textit{selective mixing}). \Fref{fig:ProbAmps} shows the amplitude probabilities of all states grouped in bins as a function of their position at the interferometer output. The two probabilities which correspond to the main pattern are in the highest bin, all the other bins correspond to the parasitic pattern. The total number of states is about 1100000.

Since the path and interactions of the main pattern are known, it is possible to determine its probability amplitude in absence of interference by manually selecting the elements of $\Tilde{L}$ and $\hat{P}$ belonging to this path. By comparing this result with the selective mixing it is possible to verify that they coincide, proving that the parasitic pattern does not interfere with the main one. Since the interaction between states depends only on the geometry of the interferometer, this also applies to non-zero angles. For this reason, it was decided to resort to computationally cheaper simulations with a semi-classical approach: each atom is described by a classical point-like particle that crosses the laser beams with which it interacts in the way described by the quantum treatment of the system \eref{eqn:system}.
This semi-classical method was compared with the selective mixing, showing that their efficiencies coincide, thus validating the results of the semi-classical simulation. The semi-classical method was also used to build a Monte Carlo simulation in order to estimate the interferometric contrast necessary to determine the acquisition time of the signal. In this algorithm the interaction between atom and laser, with consequent change of state, occurs in the case in which the transition probability between g and e is greater than a random number ranging from 0 to 1. The Monte Carlo parameters of the positronium beam are the entrance angles and positions. The input coordinates have a uniform distribution on the entrance hole and the angular distributions are Gaussian.
In the semi-classical approach, the annihilation and radiative decay were also modeled by comparing the probability with a random number, like the laser-atom interaction, in every small step in which the space was divided (spatial mesh). The $n=3$ Ps (n is the principal quantum number) can radiatively decay towards $n=2$ or $n=1$ with a probability equal to $1-\rme^{-t/\tau}$, with $\tau_{3\rightarrow 2}\simeq$89 ns and $\tau_{3\rightarrow 1}\simeq$12 ns \cite{PhysRevA.100.063414} while the $n=2$ state radiatively decays into $n=1$ in 0.24 s and annihilates in 1.13 $\mu$s. It may happen that a Ps decays radiatively from $n=3$ to $n=2$ and then continues to run through the interferometer remaining in resonance with the laser excitation. In this case the atom loses the gravitational information, and it therefore adds noise on the detector. An atom that reaches the level with $n=1$ continues to propagate until it decays or hits the detector, contributing to the noise. Each radiative decay involves the isotropic emission of a photon that randomly changes the atomic momentum.

\section*{References}
\bibliographystyle{unsrt}
\bibliography{Bibliography}

\begin{thebibliography}{10}

\bibitem{PhysRev.82.455}
M.~Deutsch.
\newblock Evidence for the formation of positronium in gases.
\newblock {\em Phys. Rev.}, 82:455--456, May 1951.

\bibitem{RevModPhys.53.127}
A.~Rich.
\newblock Recent experimental advances in positronium research.
\newblock {\em Rev. Mod. Phys.}, 53:127--165, Jan 1981.

\bibitem{LIANG1988419}
E.~Liang and C.~D. Charles.
\newblock Laser cooling of positronium.
\newblock {\em Optics Communications}, 65(6):419--424, 1988.

\bibitem{KARSHENBOIM2004}
G.~K. Savely.
\newblock Precision study of positronium: Testing bound state {QED} theory.
\newblock {\em International Journal of Modern Physics A}, 19(23):3879--3896,
  2004.

\bibitem{KARSHENBOIM20051}
G.~K. Savely.
\newblock Precision physics of simple atoms: {QED} tests, nuclear structure and
  fundamental constants.
\newblock {\em Physics Reports}, 422(1):1--63, 2005.

\bibitem{GOVAERTS1996451}
J.~Govaerts and M.~V. Caillie.
\newblock Neutrino decay of positronium in the {Standard Model} and beyond.
\newblock {\em Physics Letters B}, 381(4):451--457, 1996.

\bibitem{Castelli2012}
F.~Castelli.
\newblock The positronium atom as a benchmark for {Rydberg} excitation
  experiments in atomic physics.
\newblock {\em The European Physical Journal Special Topics}, 203:137--150, Apr
  2012.

\bibitem{PhysRevA.40.5526}
C.~D. Dermer and J.~C. Weisheit.
\newblock Perturbative analysis of simultaneous {Stark} and {Zeeman} effects on
  n=1-n=2 radiative transitions in positronium.
\newblock {\em Phys. Rev. A}, 40:5526--5532, Nov 1989.

\bibitem{PhysRevA.7.447}
S.~M. Curry.
\newblock Combined {Zeeman} and motional {Stark} effects in the first excited
  state of positronium.
\newblock {\em Phys. Rev. A}, 7:447--450, Feb 1973.

\bibitem{PhysRevA.8.625}
M.~L. Lewis and V.~W. Hughes.
\newblock Higher-order relativistic contributions to the combined {Zeeman} and
  motional {Stark} effects in positronium.
\newblock {\em Phys. Rev. A}, 8:625--639, Aug 1973.

\bibitem{e-Fall}
F.~C. Witteborn and W.~M. Fairbank.
\newblock Experimental comparison of the gravitational force on freely falling
  electrons and metallic electrons.
\newblock {\em Phys. Rev. Lett.}, 19:1049--1052, Oct 1967.

\bibitem{Cassidy2018}
D.~B. Cassidy.
\newblock Experimental progress in positronium laser physics.
\newblock {\em The European Physical Journal D}, 72:53, Mar 2018.

\bibitem{RevModPhys.29.423}
H.~Bondi.
\newblock Negative mass in general relativity.
\newblock {\em Rev. Mod. Phys.}, 29:423--428, Jul 1957.

\bibitem{doi:10.1119/1.1996159}
P.~Morrison.
\newblock Approximate nature of physical symmetries.
\newblock {\em American Journal of Physics}, 26(6):358--368, 1958.

\bibitem{SCHERK1979265}
J.~Scherk.
\newblock Antigravity: A crazy idea?
\newblock {\em Physics Letters B}, 88(3):265--267, 1979.

\bibitem{NIETO1991221}
M.~M. Nieto and T.~Goldman.
\newblock The arguments against “antigravity” and the gravitational
  acceleration of antimatter.
\newblock {\em Physics Reports}, 205(5):221--281, 1991.

\bibitem{PhysRevLett.66.850}
E.~G. Adelberger, B.~R. Heckel, C.~W. Stubbs, and Y.~Su.
\newblock Does antimatter fall with the same acceleration as ordinary matter?
\newblock {\em Phys. Rev. Lett.}, 66:850--853, Feb 1991.

\bibitem{HUBER20001245}
F.~M. Huber, R.~A. Lewis, E.~W. Messerschmid, and G.~A. Smith.
\newblock Precision tests of {Einstein's Weak Equivalence Principle} for
  antimatter.
\newblock {\em Advances in Space Research}, 25(6):1245--1249, 2000.

\bibitem{Chardin2018}
G.~Chardin and G.~Manfredi.
\newblock Gravity, antimatter and the {Dirac-Milne} universe.
\newblock {\em Hyperfine Interactions}, 239(1):45, Oct 2018.

\bibitem{Rousselle2022}
O.~Rousselle, P.~Clad{\'e}, S.~Guellati-Kh{\'e}lifa, R.~Gu{\'e}rout, and
  S.~Reynaud.
\newblock Quantum interference measurement of the free fall of anti-hydrogen.
\newblock {\em The European Physical Journal D}, 76(11):209, Nov 2022.

\bibitem{OBERTHALER2002129}
M.~K. Oberthaler.
\newblock Anti-matter wave interferometry with positronium.
\newblock {\em Nuclear Instruments and Methods in Physics Research Section B:
  Beam Interactions with Materials and Atoms}, 192(1):129--134, 2002.

\bibitem{Mariazzi2020}
S.~Mariazzi, R.~Caravita, M.~Doser, G.~Nebbia, and R.~S. Brusa.
\newblock Toward inertial sensing with a {2$^3$S} positronium beam.
\newblock {\em The European Physical Journal D}, 74(4):79, Apr 2020.

\bibitem{Zimmer21}
P.~Crivelli, D.~A. Cooke, and S.~Friedreich.
\newblock Experimental considerations for testing antimatter antigravity using
  positronium 1s-2s spectroscopy.
\newblock {\em International Journal of Modern Physics: Conference Series},
  30:1460257, 2014.

\bibitem{Cassidy2014}
D.~B. Cassidy and S.~D. Hogan.
\newblock Atom control and gravity measurements using {Rydberg} positronium.
\newblock {\em International Journal of Modern Physics: Conference Series},
  30:1460259, Jan 2014.

\bibitem{Amole2013}
C.~Amole~et al.
\newblock Description and first application of a new technique to measure the
  gravitational mass of antihydrogen.
\newblock {\em Nature Communications}, 4(1):1785, Apr 2013.

\bibitem{PhysRevLett.129.173204}
C.~Solaro, C.~Debavelaere, P.~Clad\'e, and S.~Guellati-Khelifa.
\newblock Atom interferometer driven by a picosecond frequency comb.
\newblock {\em Phys. Rev. Lett.}, 129:173204, Oct 2022.

\bibitem{MILLS2002102}
A.P. Mills and M.~Leventhal.
\newblock Can we measure the gravitational free fall of cold rydberg state
  positronium?
\newblock {\em Nuclear Instruments and Methods in Physics Research Section B:
  Beam Interactions with Materials and Atoms}, 192(1):102--106, 2002.

\bibitem{PhysRevA.94.012507}
S.~Aghion~et al.
\newblock Laser excitation of the $n=3$ level of positronium for antihydrogen
  production.
\newblock {\em Phys. Rev. A}, 94:012507, Jul 2016.

\bibitem{Mukhanov}
V.~Mukhanov.
\newblock Physical foundations of cosmology.
\newblock {\em Cambridge University Press}, 2005:24, 2005.

\bibitem{Sakharov}
A.~D. Sakharov.
\newblock Violation of {$CP$} invariance, {$C$} asymmetry, and baryon asymmetry
  of the universe.
\newblock {\em JETP Lett.}, 5:24, 1967.

\bibitem{PhysRevLett.89.231602}
O.~W. Greenberg.
\newblock {$CPT$} violation implies violation of {Lorentz} invariance.
\newblock {\em Phys. Rev. Lett.}, 89:231602, Nov 2002.

\bibitem{PhysRevD.55.6760}
D.~Colladay and V.~A. Kosteleck\'y.
\newblock $\mathrm{CPT}$ violation and the {Standard Model}.
\newblock {\em Phys. Rev. D}, 55:6760--6774, Jun 1997.

\bibitem{PhysRevD.58.116002}
D.~Colladay and V.~A. Kosteleck\'y.
\newblock {Lorentz}-violating extension of the {Standard Model}.
\newblock {\em Phys. Rev. D}, 58:116002, Oct 1998.

\bibitem{PhysRevD.92.056002}
V.~A. Kosteleck\'y and A.~J. Vargas.
\newblock {Lorentz} and {$CPT$} tests with hydrogen, antihydrogen, and related
  systems.
\newblock {\em Phys. Rev. D}, 92:056002, Sep 2015.

\bibitem{RevModPhys.81.1051}
A.~D. Cronin, J.~Schmiedmayer, and D.~E. Pritchard.
\newblock Optics and interferometry with atoms and molecules.
\newblock {\em Rev. Mod. Phys.}, 81:1051--1129, Jul 2009.

\bibitem{APeters_2001}
A.~Peters, K.~Y. Chung, and S.~Chu.
\newblock High-precision gravity measurements using atom interferometry.
\newblock {\em Metrologia}, 38(1):25, Feb 2001.

\bibitem{PhysRevA.65.033608}
J.~M. McGuirk, G.~T. Foster, J.~B. Fixler, M.~J. Snadden, and M.~A. Kasevich.
\newblock Sensitive absolute-gravity gradiometry using atom interferometry.
\newblock {\em Phys. Rev. A}, 65:033608, Feb 2002.

\bibitem{PhysRevLett.97.240801}
D.~S. Durfee, Y.~K. Shaham, and M.~A. Kasevich.
\newblock Long-term stability of an area-reversible atom-interferometer
  {Sagnac} gyroscope.
\newblock {\em Phys. Rev. Lett.}, 97:240801, Dec 2006.

\bibitem{doi:10.1126/science.aap7706}
R.~H. Parker, C.~Yu, W.~Zhong, B.~Estey, and H.~Müller.
\newblock Measurement of the fine-structure constant as a test of the {Standard
  Model}.
\newblock {\em Science}, 360(6385):191--195, 2018.

\bibitem{Rosi2014}
G.~Rosi, F.~Sorrentino, L.~Cacciapuoti, M.~Prevedelli, and G.~M. Tino.
\newblock Precision measurement of the {Newtonian} gravitational constant using
  cold atoms.
\newblock {\em Nature}, 510(7506):518--521, Jun 2014.

\bibitem{Tino2021}
G.~M. Tino.
\newblock Testing gravity with cold atom interferometry: results and prospects.
\newblock {\em Quantum Science and Technology}, 6(2):024014, Mar 2021.

\bibitem{PhysRevLett.67.181}
M.~Kasevich and S.~Chu.
\newblock Atomic interferometry using stimulated {Raman} transitions.
\newblock {\em Phys. Rev. Lett.}, 67:181--184, Jul 1991.

\bibitem{PhysRevLett.75.2638}
D.~M. Giltner, R.~W. McGowan, and S.~A. Lee.
\newblock Atom interferometer based on {Bragg} scattering from standing light
  waves.
\newblock {\em Phys. Rev. Lett.}, 75:2638--2641, Oct 1995.

\bibitem{BORDE198910}
C.~J. Bordé.
\newblock Atomic interferometry with internal state labelling.
\newblock {\em Physics Letters A}, 140(1):10--12, 1989.

\bibitem{PhysRevLett.85.4498}
J.~M. McGuirk, M.~J. Snadden, and M.~A. Kasevich.
\newblock Large area light-pulse atom interferometry.
\newblock {\em Phys. Rev. Lett.}, 85:4498--4501, Nov 2000.

\bibitem{PhysRevLett.100.180405}
H.~M\"uller, S.~Chiow, Q.~Long, S.~Herrmann, and S.~Chu.
\newblock Atom interferometry with up to 24-photon-momentum-transfer beam
  splitters.
\newblock {\em Phys. Rev. Lett.}, 100:180405, May 2008.

\bibitem{PhysRevLett.79.784}
S.~B. Cahn, A.~Kumarakrishnan, U.~Shim, T.~Sleator, P.~R. Berman, and
  B.~Dubetsky.
\newblock Time-domain de {Broglie} wave interferometry.
\newblock {\em Phys. Rev. Lett.}, 79:784--787, Aug 1997.

\bibitem{PhysRevLett.73.2563}
M.~Weitz, B.~C. Young, and S.~Chu.
\newblock Atomic interferometer based on adiabatic population transfer.
\newblock {\em Phys. Rev. Lett.}, 73:2563--2566, Nov 1994.

\bibitem{Wicht_2002}
A.~Wicht, J.~M. Hensley, E.~Sarajlic, and S.~Chu.
\newblock A preliminary measurement of the fine structure constant based on
  atom interferometry.
\newblock {\em Physica Scripta}, 2002(T102):82, Jan 2002.

\bibitem{PhysRevD.97.075020}
A.~Arvanitaki, P.~W. Graham, J.~M. Hogan, S.~Rajendran, and K.~Van~Tilburg.
\newblock Search for light scalar dark matter with atomic gravitational wave
  detectors.
\newblock {\em Phys. Rev. D}, 97:075020, Apr 2018.

\bibitem{El-Neaj2020}
Y.~A. El-Neaj.
\newblock {AEDGE}: Atomic experiment for dark matter and gravity exploration in
  space.
\newblock {\em EPJ Quantum Technology}, 7(1):6, Mar 2020.

\bibitem{PhysRevD.78.122002}
S.~Dimopoulos, P.~W. Graham, J.~M. Hogan, M.~A. Kasevich, and S.~Rajendran.
\newblock Atomic gravitational wave interferometric sensor.
\newblock {\em Phys. Rev. D}, 78:122002, Dec 2008.

\bibitem{Yu2011}
N.~Yu and M.~Tinto.
\newblock Gravitational wave detection with single-laser atom interferometers.
\newblock {\em General Relativity and Gravitation}, 43(7):1943--1952, Jul 2011.

\bibitem{PhysRevD.93.021101}
W.~Chaibi, R.~Geiger, B.~Canuel, A.~Bertoldi, A.~Landragin, and P.~Bouyer.
\newblock Low frequency gravitational wave detection with ground-based atom
  interferometer arrays.
\newblock {\em Phys. Rev. D}, 93:021101, Jan 2016.

\bibitem{Loriani_2019}
S.~Loriani, D.~Schlippert, C.~Schubert, S.~Abend, H.~Ahlers, W.~Ertmer,
  J.~Rudolph, J.~M. Hogan, M.~A. Kasevich, E.~M. Rasel, and N.~Gaaloul.
\newblock Atomic source selection in space-borne gravitational wave detection.
\newblock {\em New Journal of Physics}, 21(6):063030, Jun 2019.

\bibitem{doi:10.1142/S0218271819400054}
Ming-Sheng Zhan~et al.
\newblock {ZAIGA}: {Zhaoshan} long-baseline atom interferometer gravitation
  antenna.
\newblock {\em International Journal of Modern Physics D}, 29(04):1940005,
  2020.

\bibitem{Schubert2019}
C.~Schubert, D.~Schlippert, S.~Abend, E.~Giese, A.~Roura, W.~P. Schleich,
  W.~Ertmer, and E.~M. Rasel.
\newblock Scalable, symmetric atom interferometer for infrasound gravitational
  wave detection, 2019.

\bibitem{PhysRevLett.124.083604}
J.~Rudolph, T.~Wilkason, M.~Nantel, H.~Swan, C.~M. Holland, Y.~Jiang, B.~E.
  Garber, S.~P. Carman, and J.~M. Hogan.
\newblock Large momentum transfer clock atom interferometry on the 689 nm
  intercombination line of strontium.
\newblock {\em Phys. Rev. Lett.}, 124:083604, Feb 2020.

\bibitem{Graham2017}
P.~W. Graham, J.~M. Hogan, M.~A. Kasevich, S.~Rajendran, and R.~W. Romani.
\newblock Mid-band gravitational wave detection with precision atomic sensors,
  2017.

\bibitem{PhysRevLett.110.171102}
P.~W. Graham, J.~M. Hogan, M.~A. Kasevich, and S.~Rajendran.
\newblock New method for gravitational wave detection with atomic sensors.
\newblock {\em Phys. Rev. Lett.}, 110:171102, Apr 2013.

\bibitem{BATELAAN199785}
H.~Batelaan, S.~Bernet, M.~K. Oberthaler, E.~M. Rasel, J.~Schmiedmayer, and
  A.~Zeilinger.
\newblock - classical and quantum atom fringes.
\newblock In Paul~R. Berman, editor, {\em Atom Interferometry}, pages 85--120.
  Academic Press, San Diego, 1997.

\bibitem{Nagashima_2006}
Y.~Nagashima and T.~Sakai.
\newblock First observation of positronium negative ions emitted from tungsten
  surfaces.
\newblock {\em New Journal of Physics}, 8(12):319, Dec 2006.

\bibitem{NAGASHIMA201495}
Y.~Nagashima.
\newblock Experiments on positronium negative ions.
\newblock {\em Physics Reports}, 545(3):95--123, 2014.

\bibitem{Michishio}
K.~Michishio, T.~Kanai, S.~Kuma, T.~Azuma, K.~Wada, I.~Mochizuki, T.~Hyodo,
  A.~Yagishita, and Y.~Nagashima.
\newblock Observation of a shape resonance of the positronium negative ion.
\newblock {\em Nature Communications}, 7(1):11060, Mar 2016.

\bibitem{Consolati2013}
G.~Consolati, R.~Ferragut, A.~Galarneau, F.~Di~Renzo, and F.~Quasso.
\newblock Mesoporous materials for antihydrogen production.
\newblock {\em Chem. Soc. Rev.}, 42:3821--3832, 2013.

\bibitem{PhysRevA.98.013402}
S.~Aghion~et al.
\newblock Producing long-lived $2{\phantom{\rule{0.16em}{0ex}}}^{3}s$
  positronium via $3{\phantom{\rule{0.16em}{0ex}}}^{3}p$ laser excitation in
  magnetic and electric fields.
\newblock {\em Phys. Rev. A}, 98:013402, Jul 2018.

\bibitem{PhysRevA.81.052703}
P.~Crivelli, U.~Gendotti, A.~Rubbia, L.~Liszkay, P.~Perez, and C.~Corbel.
\newblock Measurement of the orthopositronium confinement energy in mesoporous
  thin films.
\newblock {\em Phys. Rev. A}, 81:052703, May 2010.

\bibitem{Nagashima_2021}
Y.~Nagashima, K.~Michishio, L.~Chiari, and Y.~Nagata.
\newblock An energy-tunable positronium beam produced via photodetachment of
  positronium negative ions and its applications.
\newblock {\em Journal of Physics B: Atomic, Molecular and Optical Physics},
  54(21):212001, dec 2021.

\bibitem{Sacerdoti}
M.~Sacerdoti, V.~Toso, G.~Vinelli, G.~Rosi, L.~Salvi, G.~M. Tino,
  M.~Giammarchi, and R.~Ferragut.
\newblock Towards the formation of a positronium coherent beam.
\newblock {\em arXiv:2307.12894}, 2023.

\bibitem{Igarashi_2000}
Akinori Igarashi, Isao Shimamura, and Nobuyuki Toshima.
\newblock Photodetachment cross sections of the positronium negative ion.
\newblock {\em New Journal of Physics}, 2(1):17, Aug 2000.

\bibitem{Brixino}
A.~Bacci~et al.
\newblock {$TDR$} {$BriXSinO$}:
  \url{https://marix.mi.infn.it/wp-content/uploads/2022/04/BriXSino_TDR.pdf}.

\bibitem{Drebot:2022nho}
I.~Drebot~et al.
\newblock {{$BriXSinO$} high-flux dual X-ray and THz radiation source based on
  energy recovery linacs}.
\newblock {\em JACoW}, IPAC2022:THOXSP2, 2022.

\bibitem{Bacci}
A.~Bacci, F.~Broggi, V.~Petrillo, and L.~Serafini.
\newblock Low emittance positron beam generation: a comparison between
  photo-production and electro-production.
\newblock {\em arXiv:2103.13167v1}, 2021.

\bibitem{CHARLTON2021164657}
M.~Charlton~et al.
\newblock Positron production using a 9 mev electron linac for the {$GBAR$}
  experiment.
\newblock {\em Nuclear Instruments and Methods in Physics Research Section A:
  Accelerators, Spectrometers, Detectors and Associated Equipment}, 985:164657,
  2021.

\bibitem{Cooper1968}
J.~Cooper and R.~N. Zare.
\newblock Angular distribution of photoelectrons.
\newblock {\em The Journal of Chemical Physics}, 48:942--943, 1 1968.

\bibitem{Michishio2019}
Michishio K., L.~Chiari, F.~Tanaka, N.~Oshima, and Y.~Nagashima.
\newblock A high-quality and energy-tunable positronium beam system employing a
  trap-based positron beam.
\newblock {\em Review of Scientific Instruments}, 90, 2 2019.

\bibitem{PhysRevLett.125.073002}
L.~Gurung, T.~J. Babij, S.~D. Hogan, and D.~B. Cassidy.
\newblock Precision microwave spectroscopy of the positronium $n=2$ fine
  structure.
\newblock {\em Phys. Rev. Lett.}, 125:073002, Aug 2020.

\bibitem{PhysRevA.100.063414}
M.~Antonello~et al.
\newblock Efficient {2$^3$S} positronium production by stimulated decay from
  the {3$^3$P} level.
\newblock {\em Phys. Rev. A}, 100:063414, Dec 2019.

\bibitem{PhysRevA.41.3478}
G.~Feinberg, A.~Rich, and J.~Sucher.
\newblock Quadratic zeeman effect in positronium.
\newblock {\em Phys. Rev. A}, 41:3478--3480, Apr 1990.

\bibitem{10.1117/12.907914}
O.~Homburg and T.~Mitra.
\newblock {Gaussian-to-top-hat beam shaping: an overview of parameters,
  methods, and applications}.
\newblock In Alexis~V. Kudryashov, Alan~H. Paxton, and Vladimir~S. Ilchenko,
  editors, {\em Laser Resonators, Microresonators, and Beam Control XIV},
  volume 8236, page 82360A. International Society for Optics and Photonics,
  SPIE, 2012.

\bibitem{10.1117/12.930284}
A.~Laskin and V.~Laskin.
\newblock {Imaging techniques with refractive beam shaping optics}.
\newblock In Andrew Forbes and Todd~E. Lizotte, editors, {\em Laser Beam
  Shaping XIII}, volume 8490, page 84900J. International Society for Optics and
  Photonics, SPIE, 2012.

\bibitem{Ngcobo:13}
S.~Ngcobo, K.~Ait-Ameur, I.~Litvin, A.~Hasnaoui, and A.~Forbes.
\newblock Tuneable {Gaussian} to flat-top resonator by amplitude beam shaping.
\newblock {\em Opt. Express}, 21(18):21113--21118, Sep 2013.

\bibitem{Vinelli_2020}
G.~Vinelli, R.~Ferragut, M.~Giammarchi, G.~Maero, M.~Romé, and V.~Toso.
\newblock Real-time monitoring of a positron beam using a microchannel plate in
  single-particle mode.
\newblock {\em Journal of Instrumentation}, 15(11):P11030, Nov 2020.

\bibitem{doi:10.1063/1.5135842}
C.~N. Taylor, T.~F. Fuerst, and M.~Shimada.
\newblock Characterization of coincidence {Doppler} broadening and positron
  annihilation lifetime systems at inl.
\newblock {\em AIP Conference Proceedings}, 2182(1):040010, 2019.

\bibitem{doi:10.1063/1.123815}
M.~P. Petkov, M.~H. Weber, K.~G. Lynn, K.~P. Rodbell, and S.~A. Cohen.
\newblock {Doppler} broadening positron annihilation spectroscopy: A technique
  for measuring open-volume defects in silsesquioxane spin-on glass films.
\newblock {\em Applied Physics Letters}, 74(15):2146--2148, 1999.

\bibitem{Rosi2017}
G.~Rosi, G.~D'Amico, L.~Cacciapuoti, F.~Sorrentino, M.~Prevedelli, M.~Zych,
  {\v{C}}~Brukner, and G.~M. Tino.
\newblock Quantum test of the equivalence principle for atoms in coherent
  superposition of internal energy states.
\newblock {\em Nature Communications}, 8(1):15529, Jun 2017.

\bibitem{PhysRevA.89.023607}
F.~Sorrentino, Q.~Bodart, L.~Cacciapuoti, Y.-H. Lien, M.~Prevedelli, G.~Rosi,
  L.~Salvi, and G.~M. Tino.
\newblock Sensitivity limits of a {Raman} atom interferometer as a gravity
  gradiometer.
\newblock {\em Phys. Rev. A}, 89:023607, Feb 2014.

\bibitem{Foster:02}
G.~T. Foster, J.~B. Fixler, J.~M. McGuirk, and M.~A. Kasevich.
\newblock Method of phase extraction between coupled atom interferometers using
  ellipse-specific fitting.
\newblock {\em Opt. Lett.}, 27(11):951--953, Jun 2002.

\bibitem{Chiow:12}
S.~Chiow, T.~Kovachy, J.~M. Hogan, and M.~A. Kasevich.
\newblock Generation of 43 w of quasi-continuous 780 nm laser light via
  high-efficiency, single-pass frequency doubling in periodically poled lithium
  niobate crystals.
\newblock {\em Opt. Lett.}, 37(18):3861--3863, Sep 2012.

\bibitem{Seise:10}
E.~Seise, A.~Klenke, J.~Limpert, and A.~T\"{u}nnermann.
\newblock Coherent addition of fiber-amplified ultrashort laser pulses.
\newblock {\em Opt. Express}, 18(26):27827--27835, Dec 2010.

\bibitem{Meshkov}
I.~N. Meshkov, V.~N. Pavlov, A.~O. Sidorin, and S.~L. Yakovenko.
\newblock A cryogenic source of slow monochromatic positrons.
\newblock {\em Instruments and Experimental Techniques}, 50(5):639--645, Sep
  2007.

\bibitem{TrezziAegisPhD}
D.~Trezzi.
\newblock {\em Study of Positronium Converters in the {AEgIS} Antimatter
  Experiment}.
\newblock {PhD} thesis, Università degli Studi di Milano, 2010.

\bibitem{PhysRevA.77.023609}
H.~M\"uller, S.~Chiow, and S.~Chu.
\newblock Atom-wave diffraction between the {Raman}-{Nath} and the {Bragg}
  regime: Effective {Rabi} frequency, losses, and phase shifts.
\newblock {\em Phys. Rev. A}, 77:023609, Feb 2008.

\bibitem{NOTA}
We can define the spatial extension of the Ps wave function in the laser beams
  propagation direction as $\Delta y_{\rm Ps}=\hbar/\Delta p$ and $\Delta
  p=2m_\rme\Delta v_y=2m_\rme v_ztg(\Delta\theta_y)$, where m$_\rme$ is the
  mass of the electron, v$_{\rm z}$ is the Ps velocity in the direction of
  propagation assumed constant and $\theta y$ is the Ps entrance angle between
  the directions z and y. We can assume a collimation that sets $\Delta\theta
  y=1$ mrad, which gives $\Delta y_{\rm Ps}\simeq$ 10 nm. The minimum spatial
  separation between the centres of two wave functions that propagate in the
  interferometer is $\Delta y_{\rm min}=\frac{2\Delta z\hbar k}{v_z2m_e}\simeq$
  0.38$\mu$m with the distance $\Delta z$ between two laser beams equal to 4
  mm. Given that $\Delta y_{\rm Ps}<\Delta y_{\rm min}$ we can consider that
  two wave functions overlap only when their trajectories intersect at a laser
  pulse.

\end{thebibliography}

\end{document}